# PRESOLAR GRAINS FROM NOVAE: EVIDENCE FROM NEON AND HELIUM ISOTOPES IN COMET DUST COLLECTIONS

ROBERT O. PEPIN[1], RUSSELL L. PALMA[1,2], ROBERT D. GEHRZ[3] AND SUMNER STARRFIELD[4]

[1]Department of Physics, University of Minnesota, Minneapolis, Minnesota 55455, USA; pepin001@umn.edu

[2]Department of Physics and Astronomy, Minnesota State University, Mankato, Minnesota 56001, USA

[3]Department of Astronomy, University of Minnesota, Minneapolis, Minnesota 55455, USA

[4]School of Earth and Space Exploration, Arizona State University, Tempe, Arizona 85287, USA

## ABSTRACT

Presolar grains in meteorites and interplanetary dust particles (IDPs) carry non-solar isotopic signatures pointing to origins in supernovae, giant stars, and possibly other stellar sources. There have been suggestions that some of these grains condensed in the ejecta of classical nova outbursts, but the evidence is ambiguous. We report neon and helium compositions in particles captured on stratospheric collectors flown to sample materials from comets 26P/Grigg-Skjellerup and 55P/Tempel-Tuttle that point to condensation of their gas carriers in the ejecta of a neon (ONe) nova. The absence of detectable $^{3}$He in these particles indicates space exposure to solar wind (SW) irradiation of a few decades at most, consistent with origins in cometary dust streams. Measured $^{4}$He/$^{20}$Ne, $^{20}$Ne/$^{22}$Ne, $^{21}$Ne/$^{22}$Ne and $^{20}$Ne/$^{21}$Ne isotope ratios, and a low upper limit on $^{3}$He/$^{4}$He, are in accord with calculations of nucleosynthesis in neon nova outbursts. Of these, the uniquely low $^{4}$He/$^{20}$Ne and high $^{20}$Ne/$^{22}$Ne ratios are the most diagnostic, reflecting the large predicted $^{20}$Ne abundances in the ejecta of such novae. The correspondence of measured Ne and He compositions in cometary matter with theoretical predictions is evidence for the presence of presolar grains from novae in the early solar system.

## 1. INTRODUCTION

Discoveries of interstellar grains in meteorites and interplanetary dust particles (IDPs) have added powerful observational constraints to astrophysical nucleosynthesis theory. Laboratory measurements of isotopic distributions in these presolar grains can directly constrain theoretical predictions derived from the nuclear reaction networks used in stellar evolution calculations. This connection has catalyzed a vigorous and expanding experimental effort to locate and characterize presolar grains in solar system materials and to identify their stellar sources. Detailed studies of presolar diamond, silicon carbide, graphite, silicates, oxides and other minerals have revealed isotopic distributions consistent with nucleosynthesis in supernova explosions and expanding envelopes of red giant and asymptotic giant branch (AGB) stars (Anders & Zinner 1993; Nittler 2003; Zinner 2004).

Other types of stellar outbursts or mass loss from evolved stars are also likely contributors to the presolar grain population. Recent theoretical and observational results of studies of nucleosynthesis and dust production in classical novae (e.g., Bode & Evans 2008; José et al. 2004) have prompted interest in the possibility of nova sources. Classical nova explosions on white dwarf stars are frequent, appearing in the present galaxy at a rate of ~30-40 per year; approximately 25-33% of these occur in systems containing Ne-rich white dwarfs and have been named “neon novae” (Gehrz et al. 1985, 1998). There is suspicion, but no consensus, that isotopic distributions in some presolar grains may originate in nova outbursts. Neon nova sources for a few silicon carbide and graphite particles in the Murchison meteorite were proposed on the basis of C, N, and Si isotopic measurements (Amari et al. 2001; José et al. 2004) but there are counter arguments favoring a supernova origin (Nittler & Hoppe 2005). Another Murchison SiC grain with extreme C and N isotope ratios could be a nova candidate (Nittler et al. 2006).

The He and Ne data reported here display isotopic signatures of materials from a neon nova in IDPs from comets. Similar noble gas analyses of bulk meteorite samples in the 1960s revealed the first isotopic evidence for presolar grains in solar system matter, those carrying the exotic Xe (Reynolds & Turner 1964) and Ne (Black & Pepin 1969; Black 1972) components now known as Xe-HL and Ne-E. At that time the gas-carrying particles in these samples had not yet been identified. Subsequent experimental and theoretical work succeeded in separating the rare carrier grains from the bulk meteorites and relating their gas compositions to nuclear processing in particular types of stars (Anders & Zinner 1993; Zinner 2004). The situation here is comparable to those early studies in that the mineralogies and sizes of the carrier grains hosting the anomalous gases found in some of our samples are not yet established.

Modern searches for interstellar grains in meteorites and IDPs are carried out for the most part by scanning samples for small "hotspots" of isotopically anomalous H, C, N, O, Mg, Si and other elements utilizing secondary ion mass spectrometry (SIMS), scanning electron microscopy (SEM), and other analytic techniques with high spatial resolution (e.g., Amari et al 2001; Busemann et al. 2009). Instruments of this type, however, cannot detect trace amounts of noble gases. Measurement of these elements by static mass spectrometry requires samples considerably larger than the individual sub-micrometer (μm) presolar grains typically found in hotspot searches. The Ne and He data reported here were obtained from ~8-15μm bulk fragments of IDPs in stratospheric collections that maximized the chances of capturing infalling cometary material (Messenger 2002, Zolensky 2004). Samples from these collections are unusually rich in presolar grains (Busemann et al. 2009).

## 2. SAMPLES AND EXPERIMENTAL PROCEDURES

### *2.1. Sample Preparation and Analysis*

Samples were selected from the Grigg-Skjellerup and Tempel-Tuttle stratospheric collection surfaces L2054, L2055, U2121 and U2108 at the NASA Johnson Space Center Cosmic Dust Curatorial Facility (Zolensky 2004), and were packaged there in individual platinum foils for mounting in the Minnesota sample furnace. All samples are ~nanogram (ng)-mass fragments of cluster IDPs —larger particles that shattered on collision with the collection surfaces. Two separate particle allocations were analyzed for He and Ne by static mass spectrometry, the primary sampling (Suite 1, Table 1) in 2004 and a later re-sampling of the Grigg-Skjellerup collectors L2054, L2055 and U2121 (Suite 2, Table 3) as received over the period 2007-2010. Sample identifiers in Tables 1-3 were assigned by a convention in which an IDP, for example L2054 A1*1 in Suite 1, is uniquely designated first by collector (L2054), then by particle (A1), and lastly by cluster number (1). Thus L2054 A1*1 and the Suite 2 sample L2054 L4*1 are two different particles (A1 and L4) from cluster 1 on collector plate L2054. Of the 15 different clusters sampled for Suite 1, eight were re-sampled for Suite 2. The Suite 2 particles were mostly smaller in mass (0.04 to 1.2 ng) than the Suite 1 set (0.28 to 1.9 ng) and their Ne abundances were below the levels necessary for accurate isotope ratio measurements.

### *2.2. Sample Compositions*

None of the Suite 1 samples selected for noble gas analysis were mineralogically or chemically characterized before they were destructively analyzed at Minnesota by high temperature heating. Some fragments of the same cluster particles from which the Suite 1 and 2 samples were extracted, and other clusters from the Grigg-Skellerup collection, are composed primarily of silicates, aluminum oxides, GEMS (glass with embedded metal and sulfides), and carbonaceous matter in various proportions (Nakamura-Messenger et al. 2008). One of these contains the new mineral Brownleeite (Nakamura-Messenger et al. 2010). However, given the chemical and mineralogic heterogeneity observed in different fragments from a single cluster IDP (Thomas et al. 1995), the extent to which these compositional data are specifically applicable to the Suite 1 and 2 particles is uncertain.

### *2.3. Experimental Procedures and Data Acquisition*

After loading of sample and empty platinum foil packets into the multiple-sample furnace (the empty packets for blank analyses), the extraction system was baked at ~150°C for 3 days. Samples were then degassed in a series of computer-controlled 10-second heating pulses from ~350°C to a maximum temperature of ~1400°C. Helium and Ne released in successive temperature steps were accumulated in the mass spectrometer. Helium abundances were monitored during the pulse heating sequence, and heating was interrupted and evolved He and Ne abundances and isotopic compositions measured whenever enough He had accumulated for accurate isotopic measurement. Following this the spectrometer was emptied and gas accumulation resumed on initiation of the next step in the heating sequence. All sample and blank foils were heated and analyzed in the same way.

Since this procedure differs from the conventional analytic protocol in which noble gases are first purified and then released into a mass spectrometer at a single well-defined time, we describe it in more detail. The gas extraction, processing, and analysis system is composed of three sections in series: (a) a multi-sample furnace in which samples wrapped in Pt foil are connected across Pt leads and heated by stepped pulses of increasing electrical power; (b) a gas purification section equipped with Westinghouse@ WL and NP10 getters, a SAES@ metal appendage getter, and cryopumping on stainless steel (SS) and activated charcoal fingers at liquid $N_2$ temperature; and (c) a 3.75" double-focusing Mattauch-Herzog geometry mass spectrometer equipped with pulse-counting detectors (Mattauch & Herzog 1934; Nier & Schlutter 1985). Gases evolved during a heating step in (a) are released directly into section (b) where they are exposed to the getters and cryopumped for one minute on the SS finger to reduce abundances of chemically active and condensable species. After one minute the SS finger is isolated from the system and cryopumping is switched to the activated charcoal. The mass spectrometer inlet is then opened for a quick analysis of masses 4, 18, 20, 40 and 44. These masses are rapidly scanned three times in succession, and the amount of $^4He$ used to determine if the estimated threshold for accurate helium isotopic analysis has been reached. If so, heating is halted and sections (b) and (c) isolated from (a). While the evolved gases continue to be exposed to the cold charcoal finger and getters in (b), the ion source electron voltage is reduced to 35 eV (Sec. 2.6) and the mass spectrometer electronics set to maximize sensitivity for Ne. Following the Ne analysis, electron voltage is increased to 75 eV, the spectrometer re-tuned for He, and the evolved He analyzed. The spectrometer is then pumped.

If the $^4He$ released in a particular heating step is judged insufficient to provide an accurate isotopic analysis, sections (a) and (b) are again isolated from the mass spectrometer and stepped heating is resumed. Thus the gas ultimately analyzed is the sum of gases released during the number of sample heating steps required to reach either the $^4He$ threshold or the ~1400 °C maximum temperature. In samples whose heating is uninterrupted, the total time taken to complete the 14 heating steps to ~1400 °C is approximately 37 minutes, after which Ne and then He analyses are performed.

Neon analyses of samples, blanks, and calibrations comprise 10 individual blocks of computer controlled peak-to-peak jump scan measurements of masses 18, 19, 20, 22, 40, and 44 in each block. To determine the small mass 21 abundance with higher precision, it is measured 10 times individually in each of the 10 blocks, for a total of 100 measurements. Automatic re-centering on the mass 18 water signal in each block compensates for any magnetic field drift — usually negligible— that might have occurred during this time. This full Ne analysis requires approximately 60 minutes. The following ~75-minute He analysis includes jump-scan measurements of masses 2, 3, and 4, and is performed in a similar fashion. $^{4}$He is measured 10 times and $^{3}$He 100 times, with re-centerings on mass 4 (for $^{4}$He) and on mass 2 (for $^{3}$He). Since count rates for both $^{21}$Ne and $^{3}$He are generally low, signals are corrected for small electron multiplier "dark currents" measured in parallel with $^{21}$Ne and $^{3}$He.

Neon evolved in this stepped heating and data acquisition procedure may reside in the spectrometer for times as long as ~37 minutes prior to analysis, and He up to ~100 minutes, raising the possibility that isotopic distributions could have been altered during this storage period by spectrometer ion-pumping and memory contributions. Both ion pumping and memory effects generate time trends in the measurements, and are customarily removed from sample data by extrapolating measured abundances and isotope ratios vs. time back to the instant of gas inlet into the spectrometer ("time-0"). There is no definable time-0 when gases accumulate in the instrument during stepped heating, as they do here. However time trends were not observed in Ne and He sample measurements; abundances were constant within statistical uncertainties over the duration of all step-heated sample and blank analyses. This was also the case in calibrations, where no time dependences of Ne signals were seen in the ~60 minute analyses following known gas inlet times. We conclude that the absence of a time-0 marker for accumulated gases is therefore not a critical issue, and that significant perturbations of actual compositions during pre-analysis gas storage in the spectrometer are unlikely.

*2.4. Calibrations and Blanks*

Sensitivities for He ($2.58 \times 10^{-13}$) and Ne ($4.95 \times 10^{-13}$) in units of $cm^3$STP per count $s^{-1}$ (1 $cm^3$STP = $2.69 \times 10^{19}$ atoms) were determined from repeated analyses of air samples metered into the system from calibration cans designed to permit accurate volumetric splitting, with $O_2$, $N_2$ and other chemically active constituents continually purified by getters and cryopumping (Sec. 2.3). The helium isotopic standard for mass discrimination corrections was reagent grade helium spiked with $^3$He [$^3He/^4He = (7.61 \pm .08) \times 10^{-4}$] for Suite 1 and most Suite 2 IDPs. A new standard [$^3He/^4He = (5.00 \pm .05) \times 10^{-4}$] was used for the remainder of the Suite 2 samples. Air Ne isotope ratios $^{20}Ne/^{22}Ne = 9.80 \pm 0.08$, $^{21}Ne/^{22}Ne = 0.0290 \pm 0.0003$ (Eberhardt et al. 1965) served as the Ne standard.

One set of procedural blanks was measured on empty Pt foil packets in two ways: cold (14 steps, no heating) and hot (14 heating steps to ~1400°C). Averages of these cold and hot blanks within a suite of measurements were indistinguishable. There were no statistically significant differences between these values and those for re-heatings of sample foils. These observations demonstrate that the blanks derive primarily from the room-temperature gas purification system, not from degassing of the furnace hardware or Pt sample foils during stepped heating.

For Suite 1, averages of 36 He and 15 Ne blank abundances accumulated in 14 steps (in $cm^3$STP, errors ± 1σ SDOM) were $^3He = 9.8 \pm 0.3 \times 10^{-15}$, $^4He = 6.7 \pm 0.4 \times 10^{-12}$, $^{20}Ne = 8.9 \pm$ 0.4 x $10^{-13}$, $^{21}$Ne = 6.3 ± 0.4 x $10^{-15}$, and $^{22}$Ne = 8.6 ± 0.4 x $10^{-14}$. Blank isotope ratios were $^{3}$H/$^{4}$He ≃ 1.5 ± 0.1 x $10^{-3}$, $^{20}$Ne/$^{22}$Ne ≃ 10.3 ± 0.7, and $^{21}$Ne/$^{22}$Ne ≃ 7.3 ± 0.6 x $10^{-2}$. For most of Suite 2, $^{3}$He = 6.3 x $10^{-15}$, $^{4}$He = 8.6 x $10^{-12}$, $^{20}$Ne = 5.4 x $10^{-13}$ cm$^{3}$STP. Blanks were somewhat different for the last 3 Suite 2 samples: $^{3}$He = 2.6 x $10^{-15}$, $^{4}$He = 5.9 x $10^{-12}$, $^{20}$Ne = 7.9 x $10^{-13}$ cm$^{3}$STP for L2055 L1*7, and $^{3}$He = 4.9 x $10^{-15}$, $^{4}$He = 9.7 x $10^{-12}$, $^{20}$Ne = 3.1 x $10^{-12}$ cm$^{3}$STP for L2054 L3*2 and U2121 E2*2.

A separate set of procedural blanks was applied to samples where stepped heating was interrupted and interim analyses performed. Here the blank procedure duplicated that for a sample —i.e., gases accumulated through prior steps were measured at each point in the heating sequence where the sample gases had been analyzed. Again, there was no statistical difference between cold and hot blanks, nor with sample re-heatings. Blank abundances for these interim measurements were very similar to those for the uninterrupted 14-step sequence (above), independent of the number of steps to the interruption temperature. No dependence on accumulation time was observed for any blanks in this study.

Blank corrections were most critical for $^{3}$He and $^{21}$Ne (and, for samples discussed in Sec. 2.6, for $^{22}$Ne as well). A detailed analysis of the He blank sources in our extraction and analysis system was reported earlier (Pepin et al. 2000). In cases where the isotope was detectable above the blank, blank contributions to the measured gas in the cumulative sample heating ranged from 5-85% (average ~40%) for $^{3}$He and from 40-90% (average ~65%) for $^{21}$Ne. Uncertainties due to blank corrections are included in errors listed for all data in Tables 1- 3.

### *2.5 Gas Release Profiles*

The 15 IDPs in Suite 1 display two distinctly different patterns of gas release with increasing temperature. Six of the samples (#3, 4, 7, 8, 11 & 15 in Table 1) were comparatively rich in He; in five of the six, gases were measured in two or more heating steps to ~1400°C, with outgassing beginning at temperatures ranging from ~350°C to ~800°C. Blank-corrected data obtained in individual temperature steps of samples where the heating sequence was paused and gases released to that point were measured are given in Table 2. In contrast, so little He evolved from all but one of the other nine IDPs that only a single analysis, of total gases accumulated over all 14 heating steps to ~1400°C, was performed. The exception was IDP U2108 A1*1 (#6 in Table 2), which released SW-like He together with a small Ne component, isotopically resembling an air-SW mixture, in three steps between 520°C and 870°C. Little additional He was evolved in the subsequent steps to 1430°C. Analysis of this high-temperature fraction, which contained >70% of the total Ne, revealed a Ne component strikingly similar to that in the other eight low-He IDPs. U2108 A1*1 was therefore assigned to this group. The low-temperature releases could plausibly have derived from a small high-He particle adhering to this IDP, together with a minor air contaminant.

Except for U2108 A1*1 (#6) where only the 1430°C data are given, abundances, concentrations and isotope ratios reported in Tables 1 & 3 are integrations of measurements over all heating steps. It is important to emphasize that every sample, high-He and low-He, was subject to the same multi-step heating sequence for gas extraction whether or not it was interrupted for interim analyses. As was the case for most of the low-He IDPs, gases evolved from two of the high-He group were also measured only after termination of heating at maximum temperature (#15, U2108 A2*2), or close to it (#11, U2121 A3*3 —see Table 2), and therefore had relatively long pre-analysis residence times in the mass spectrometer. As seen in Table 1, they are isotopically similar to the other four high-He IDPs where interim measurements were made at lower temperatures. Note also that the sample ordering in Tables 1 & 2 is by analysis date, and that appearances of individual high- and low-He IDPs are well interspersed throughout the duration of the experimental study.

*2.6. Neon Blanks and Possible Isotopic Interferences in the Anomalous IDPs*

The nine He-poor IDPs (Sec. 2.5) are characterized by He and Ne isotope ratios that differ markedly from solar-system compositions, whereas those displayed by the six He-rich samples reflect implantation of their noble gas inventories by solar wind (SW) irradiation, in common with most IDPs (Table 1). For this and other reasons discussed in Sec. 3.1, the two groups are respectively designated as "anomalous" and "normal". Measurement of extraordinarily high $^{20}Ne/^{22}Ne$ ratios in the anomalous IDPs directed particular attention to blank and memory corrections to the low $^{22}Ne$ abundances and to the possibility of mass 20 augmentation by doubly-ionized $^{40}Ar$. The average $^{22}Ne$ blank correction was 67% with a range in individual samples of ~54-77%. Average measured $^{22}Ne$ was 50% (range ~30-85%) above the $^{22}Ne$ blank, about 10 times the blank uncertainty. Although measurement errors for $^{22}Ne$ in the samples were substantial (1σ average of ±37%, range ~ ±18-70%), minimum (-1σ) abundances were still above the blank level in 5 of the 9. Errors in Table 1 for $^{20}Ne/^{22}Ne$ ratios in the anomalous IDPs primarily reflect uncertainties in blank-corrected $^{22}Ne$ abundances.

Memory build-up during pre-analysis storage in the spectrometer is thought to be negligible (Sec. 2.3). But even if it were present, a memory Ne component with its known $^{20}Ne/^{22}Ne$ ratio of ~11 ± 1 —reflecting a residual ion-buried mixture of air-Ne and SW-Ne compositions previously analyzed in the spectrometer— would depress measured $^{20}Ne/^{22}Ne$ ratios in the low-He anomalous group below their actual values. An elevated and undetected blank contribution with $^{20}Ne/^{22}Ne = 10.3 \pm 0.7$ (Sec. 2.4) would have the same effect.

Measurements of $^{20}Ne$ generally had much smaller statistical fluctuations, and all were many standard deviations greater than the blank value. The question is whether mass 20 abundances could have been enhanced by $^{40}Ar^{++}$. Care was taken to ensure against this contribution. An electron impact voltage of 45eV was experimentally determined to be low enough to make doubly ionized $^{40}Ar$ and $^{12}C^{16}O_2$ undetectable at masses 20 and 22; for added insurance that $^{40}Ar^{++}$ and $^{12}C^{16}O_2^{++}$ were completely suppressed it was further reduced to 35eV for Ne analyses, without substantial loss of sensitivity. Moreover the spectrometer was continually cryopumped with liquid $N_2$-chilled charcoal during the ~1 hour duration of Ne analyses, reducing $^{40}Ar$ to very low levels. Plots of $^{40}Ar$ vs. $^{20}Ne$ and $CO_2$ vs. $^{22}Ne$ show no evidence of correlation.

### *2.7. Isotope Ratios Involving $^{3}He$ and $^{21}Ne$ in the Anomalous IDPs*

$^{3}He$ was not detected in any of the anomalous samples. Two of them, #13 (U2108 A3*3) and #10 (L2055 A5*8) in Table 1, contained enough $^{4}He$ that $^{3}He$ would have been seen above the $^{3}He$ blank, by factors of ~6 and ~3 times the blank uncertainty, if the $^{3}He/^{4}He$ ratio were as high as $10^{-4}$. The actual maximum is therefore lower. Although orders of magnitude higher than the theoretical —and unmeasurable— prediction of $<10^{-8}$ in a neon nova outburst (Starrfield et al. 2009), the $<10^{-4}$ upper limit is still below the $^{3}He/^{4}He$ range in solar-system matter, from $1.66 \times 10^{-4}$ in Jupiter's atmosphere (Mahaffy et al. 1998) to an average of $4.49 \times 10^{-4}$ in the SW (Table 4).

The $^{21}Ne/^{22}Ne$ and $^{20}Ne/^{21}Ne$ ratios assigned to the anomalous IDPs are from a single imprecise measurement of $^{21}Ne$ in sample #2 (L2054 A4*4). Despite the large uncertainty, minimum (-1σ) $^{20}$Ne/$^{21}$Ne is still a factor 1.7 above the SW ratio and within theoretical predictions for neon novae. Other anomalous IDPs have larger $^{20}$Ne abundances than L2054 A4*4 but contained no detectable $^{21}$Ne above blank, indicating higher $^{20}$Ne/$^{21}$Ne ratios that would be fully consistent with a neon nova source (Sec. 4.2).

## 3. RESULTS

### *3.1. Neon*

Neon concentrations in the fifteen Suite 1 IDPs from the Grigg-Sjkellerup and Tempel-Tuttle collections (Table 5) are shown in Fig. 1. Plotted concentrations are numbered to identify the corresponding IDP data in Tables 1, 2 & 5. They clearly fall into two isotopically distinct groups, one "normal" and the other "anomalous". The normal population is characterized by the presence of SW-He, and by correlated $^{20}$Ne and $^{22}$Ne concentrations with $^{20}$Ne/$^{22}$Ne close to the SW ratio of 13.9 (Table 4), identifying these as particles that experienced substantial implantation of solar ions during their residence in space. In the following discussion we define "normal" IDPs as those containing only Ne and He inventories attributable to space irradiation, by SW or by cosmic rays producing spallogenic components. None of these contain recognizable non-SW gases indigenous to the particles.

The anomalous IDPs in Fig. 1 are strikingly different. They contain no detectable SW-He, and low and approximately uniform $^{22}$Ne concentrations together with highly variable $^{20}$Ne, an unusual isotopic distribution noted in an earlier abstract (Palma et al. 2005). Measured $^{20}$Ne/$^{22}$Ne ranges from ~25 to ~150. These high values greatly exceed the SW ratio, and are not seen in meteorites, other IDPs, or planetary atmospheres. The anomalous IDPs are compared in Fig. 2 to 64 normal IDPs with measured Ne compositions, utilizing a $^{4}$He/$^{20}$Ne vs. $^{22}$Ne/$^{20}$Ne representation that allows both groups and their measurement uncertainties to be shown at high resolution. The average offset between them is evident for both $^{22}Ne/^{20}Ne$ (factor of ~6) and $^{4}He/^{20}Ne$ (factor of ~20). Within the normal IDPs themselves, the rightward scattering of $^{22}Ne/^{20}Ne$ in Fig. 2 away from the incident SW ratio toward values deficient in $^{20}Ne$ suggests varying degrees of isotopic fractionation due to sputter erosion in space (Becker 1998; Pepin et al. 1999).

The Fig. 2 data are re-plotted in Fig. 3 with Ne isotopic compositions converted to the more conventional $^{20}Ne/^{22}Ne$ representation. Collapse of the SW and normal IDP data set in Fig. 2 to a narrow band at the bottom of the diagram emphasizes the dramatically higher $^{20}Ne/^{22}Ne$ ratios in many of the anomalous IDPs.

### *3.2. Helium*

Examination of He alone confirms that the distinctly low $^{4}He/^{20}Ne$ signature in the anomalous IDPs (Fig. 2) is due to deficits in $^{4}He$ content relative to normal IDPs. In the Fig. 4 plot of $^{3}He/^{4}He$ vs. $^{4}He\ g^{-1}$ the average $^{4}He$ concentration in the anomalous particles is seen to be a factor of ~3 lower than in any of the comparison groups that collectively include 83 normal IDPs. The unique nature of this He is even more evident in $^{3}He$ contents, not normalized to particle masses. The Fig. 5 logarithmic plot of measured $^{3}He$ vs. $^{4}He$ abundances shows that the groups of normal IDPs fall closely along a power law distribution (from which an average $^{3}He/^{4}He$ ratio of ~$3.2 \times 10^{-4}$ may be derived). Sample U2108 A3*3, with the highest $^{4}He$ content of the anomalous IDPs ($2.06 \times 10^{-11}$ $cm^{3}$STP: Table 1) but no detectable $^{3}He$, sets a $^{3}He/^{4}He$ upper limit of $<10^{-4}$ (Sec. 2.7). U2108 A3*3 is clearly distinct from the normal samples, falling below their power law distribution by a factor of $\gtrsim 4$ in $^{3}He$. One of the normal groups (yellow symbol) is lower than U2108 A3*3 in average measured $^{4}He$ by about the same factor, yet SW-

$^3He$ was readily detected in all 12 of its constituent particles. The failure to see $^3He$ in U2108 A3*3 is therefore not due to a He content too low to measure a SW contribution at or well below the levels seen in normal IDPs.

*3.2.1 Space Residence Time of Anomalous IDPs in the Grigg-Skjellerup Collection*

The absence of detectable $^3He$ indicates that the anomalous IDPs have spent little time exposed to solar radiation as free particles in space. With some assumptions, an upper limit to their average space residence time in a comet dust stream orbit can be estimated. This requires, in addition to maximum $^3He$ content $[^3He]_{max}$, approximate measures of (a) the He loss factor $f_h$ from drag heating during atmospheric entry, and (b) the factor $f_s$ due to partial shielding of a cluster IDP fragment from SW irradiation by fractional burial in its larger parent particle. With the average SW-$^3He$ flux over the dust stream orbit represented as $F_3$, the upper time limit to SW exposure is $T_{max} = [^3He]_{max} f_s f_h/F_3$.

The anomalous IDPs from the Grigg-Skjellerup (GS) collectors (L2054 A4*4, L2055 A1*1, L2055 A5*8, U2121 A1*1 and U2121 A2*2 in Table 1, excluding the two with no detectable $^4He$) have an average cross-section area of ~8.0 x $10^{-7}$ $cm^2$ and a $^4He$ abundance of ~5.9 x $10^{-12}$ $cm^3$STP. With $^3He/^4He$ <$10^{-4}$, $[^3He]_{max} \simeq 2.0 \times 10^{10}$ atoms $cm^{-2}$. An average value of $F_3$ for the GS dust stream may be calculated from the near-Earth SW-$^4He$ flux $(F_4)_o = 1.24 \times 10^7$ atoms $cm^{-2}$ $s^{-1}$ (Reisenfeld et al. 2007) and the perihelion ($R_o$ ~1 AU) and aphelion (R = 4.93 AU) of the GS orbit. The average $^4He$ flux over the orbit from $R_o$ to R is $\int(F_4)_o dr/r^2/\int dr = (F_4)_o/RR_o$ = 2.51 x $10^6$ atoms $cm^{-2}$ $s^{-1}$, and the corresponding $^3He$ flux, with $^3He/^4He$ in the SW = 4.49 x $10^{-4}$ (Table 4), is $F_3 = 1.13 \times 10^3$ atoms $cm^{-2}$ $s^{-1}$.

An estimate of $f_h$ ~10 for He comes from Fig. 2 where the average $^4He/^{20}Ne$ ratio in 64 normal IDPs is seen to be approximately an order of magnitude below the SW implantation ratio of ~650 (Table 4); the assumption here is that drag heating primarily releases He while less labile Ne is largely retained. About 2/3 of the ratios in Fig. 2 fall between $^{4}He/^{20}Ne$ ~20 and ~200, corresponding to $f_h$ values in the range ~3-30. With this as an uncertainty estimate, $f_h$ is ~10 within a factor of ~3.

The SW shielding factor $f_s$ is more difficult to evaluate. Our approach was to compare $^{20}Ne$ contents (atoms $cm^{-2}$-cross section) in ~60 individual IDPs and ~40 cluster IDP fragments, utilizing data from Nier & Schlutter (1990), Pepin et al. (2000, 2001), Kehm et al. (2006), this work (Tables 1 & 3), and unpublished IDP data from the Minnesota laboratory. The comparison indicates a ~4-fold lower Ne loading in the average cluster fragment compared to the average individual particle, with a conservative range of ~1 to 7 depending on how the various data sets are weighted. This factor of 4 ± 3 can be taken as an approximate measure of $f_s$ if two implicit assumptions are roughly valid: that the individual IDPs experienced SW implantation over their entire surfaces —i.e., $f_s$ = 1 (no shielding)— and the relatively large numbers of IDPs in the individual and cluster groups have about the same average SW exposure age. A $f_s$ value of 4 means that, on average, a cluster fragment was exposed over 25% of its surface area. We note that virtually all cluster IDP fragments analyzed for noble gases, except for the anomalous ones reported here, contain $^{3}He$ attributable to SW implantation. This argues against high values for $f_s$ and implies relatively open in-space structures for the parent cluster particles, as opposed to geometries which would shield significant fractions of their volumes from short-range (~10-30 nanometer) SW radiation.

With these estimates for $f_h$ and $f_s$, the nominal upper limit for SW exposure of an average anomalous IDP originating from the GS dust stream is $T_{max}$ ≃20 years. Uncertainties in $f_h$ and $f_s$ allow ranges for $T_{max}$ of ~7-65 years and ~6-40 years respectively. The actual maximum exposure age of a dust stream IDP released after perturbation of the comet into an Earth-crossing orbit in 1964 (Messenger 2002) and collected in 2003 (Zolensky 2004) is ~40 yr. Many such particles, shed by the comet in apparitions closer in time to 2003, will have shorter exposures. The estimated average space residence time of at most a few decades for the anomalous IDPs after their assumed release from GS is therefore entirely compatible with the chronology of the GS Earth-crossing dust stream.

### *3.3. Notes on Particle Densities and Origins*

Densities have been measured for IDPs classified as asteroidal (<14 km $s^{-1}$ entry velocity) and cometary (>18 km $s^{-1}$) in origin (Joswiak et al. 2000). Classification was based on release profiles of He in laboratory stepped heating and a model of particle heating during atmospheric entry (Love & Brownlee 1994). Reported averages and ranges are 2.5 (1 – 5.4) g $cm^{-3}$ for the 12 IDPs in the asteroid group, and 1.1 (0.6 – 1.9) g $cm^{-3}$ for the 13 in the comet group. Particle masses in Tables 1 & 3, used to calculate Ne and He concentrations in Figs. 1 and 4, are estimates based on assumed densities of 2.5 g $cm^{-3}$ for the normal IDPs and 1.1 g $cm^{-3}$ for the anomalous IDPs. The wide ranges in possible densities are primarily responsible for the large uncertainties, estimated overall to be ~ ± 70%, in the masses of individual samples listed in Tables 1 & 3.

Errors in Ne concentrations ($cm^3$STP/g-IDP) listed in Table 5 and plotted in Fig. 1 do not include these large mass uncertainties; they indicate only the errors in measured $^{20}$Ne and $^{22}$Ne abundances ($cm^3$STP) given in Table 1. If mass uncertainties were included, error bars would span half the x-axis of Fig. 1 for the high $^{20}$Ne anomalous data. They are omitted because they have no impact on any of the paper's conclusions; the discussions and supporting figures, except for the anomalous $^4$He concentration plotted in Fig. 4, involve either isotope ratios, where they cancel, or abundances, where they do not enter. Consideration of mass uncertainties, however, is germane to comparisons of the concentration distributions for the normal and anomalous IDPs in Fig. 1. Use of actual masses —if we knew what they were— in calculating concentrations could generate quite different distributions.

The normal IDPs in Fig. 1 are assumed to be asteroidal. They probably belong to the background population of IDPs present on most stratospheric collectors (Zolensky 2004). Their He and Ne inventories appear to be due solely to SW irradiation, and indicate substantial time in space. All of the normal IDPs in Fig. 2 do not necessarily derive from asteroids; some could be cometary, with solar components masking any indigenous noble gas signatures that might be present. However their classification as "normal" is independent of assumed density since their $^{20}$Ne and $^{22}$Ne abundances are correlated. Different densities in the Suite 1 particles, for example, shift their positions in Fig. 1 but they remain within error on the 13.9 SW line.

There are strong arguments for assignment of cometary density to the anomalous IDPs. They are characterized by undetectable $^{3}$He, indicating minimal exposure to the SW (Sec. 3.2.1) as would be expected for recent release into cometary dust streams (Messenger 2002). They were taken from collectors unusually rich in presolar grains, consistent with expectations for primitive cometary matter (Busemann et al. 2009; Nakamura-Messenger et al. 2008, 2010), and they contain strikingly non-solar Ne and He compositions quite unlike those seen in meteorites derived from asteroid parent bodies. It seems most likely that they originated in the dust streams the collection flights were targeted to sample.

## 4. DISCUSSION

### *4.1. Stellar Source*

The absence of detectable SW components indicates that He and Ne inventories in the anomalous IDPs represent essentially unmodified indigenous gases. Of the combination of theoretical hydrodynamic calculations and nuclear reaction networks describing element synthesis in stellar environments, only those for total ejecta compositions in neon nova outbursts specifically predict the combination of negligible $^{3}$He, low $^{4}$He/$^{20}$Ne, high $^{20}$Ne/$^{21}$Ne, and $^{20}$Ne/$^{22}$Ne ratios in and above the range of ~25 to ~150 (Fig. 3) measured in these particles (Starrfield et al. 2009; José et al. 2004).

A classical nova explosion is caused by a thermonuclear runaway (TNR) on the surface of a white dwarf (WD) that is accreting matter through the inner Lagrangian point in a close binary star system (Bode & Evans 2008). Radiation pressure from the TNR ejects into space a large fraction of the accreted gas as well as core material dredged up from the WD itself during the TNR. Each nova outburst ejects material with unique abundances because of the large range in parameters affecting the properties of the explosion. These include WD mass and composition, initial WD luminosity, the mass accretion rate, and the nuclear reactions in the TNR itself (José et al. 2004; Starrfield et al. 2008, 2009). Once propelled from the WD, the expanding shell of ejected gas can be observed spectroscopically to determine its chemical composition (Gehrz et al. 1998). The discovery of a strong [Ne II] 12.8μm emission line in nova QU Vul 1984#2 provided firm observational evidence for nova eruptions driven by TNRs on oxygen-neon (ONe) white dwarf stars (Gehrz et al. 1985). Velocity broadening prevents assignment of observed Ne emission lines to a particular isotope, but the composition of the dredged-up ONe WD core material predicts that Ne initially in the ejecta is essentially all $^{20}$Ne. Detailed calculations (José et al. 2004; Starrfield et al. 2008, 2009) show that repeated proton captures on $^{20}$Ne will fuse some of the neon to $^{22}$Na. Preexisting $^{3}$He and $^{22}$Ne are virtually annihilated in the TNR by fusion to heavier species; $^3He/^4He$ is driven to $<10^{-8}$ and $^{20}Ne/^{22}Ne$ to $>10^6$ (Starrfield et al. 2009). Total $^{22}Ne$ inventories ($\Sigma^{22}Ne$) in the neon nova models are thus dominated by beta decay of $^{22}Na$ (half-life = 2.60 years), and $^3He$ is essentially absent.

Helium and Ne distributions from a recent modeling of total ejecta composition in a neon nova outburst (Starrfield et al. 2009) are given in Tables 6 & 7 and plotted as isotope ratios in Fig. 6. Their non-solar character is evident for ratios involving the enhanced $^{20}Ne$ and severely depleted $^3He$. Since modeling results are sensitive to parameter choices, Fig. 6 also shows ratio ranges derived from ten ONe WD outburst models based on differing selections for input parameters (José et al. 2004). Ratio ranges in total ejecta from supernovae, calculated from the SN II nucleosynthesis models of Rauscher et al. (2002), are plotted for comparison.

### *4.2. The Case for Origin of the Anomalous IPD Gas Carriers in a Neon Nova Outflow*

He and Ne compositions in the normal and anomalous IDPs (Table 7) are shown in Fig. 7. Solar wind ratios and calculated ratios and ranges for neon nova outbursts from Fig. 6 are plotted for comparison. Normal IDP averages fall close to SW composition, consistent with prolonged space irradiation, except for a low $^4He/^{20}Ne$ ratio suggesting loss of labile implanted He by drag heating during atmospheric entry. In contrast, as seen in Fig. 7, measured isotopic averages and ranges in the anomalous IDPs, particularly the non-solar $^4He/^{20}Ne$ and $^{20}Ne/\Sigma^{22}Ne$ ratios reflecting high $^{20}Ne$ abundances, are in good agreement with theoretical predictions for neon nova ejecta. This is also true for ratios involving $^3He$ and $^{21}Ne$, but these are less definitive since they rest on a calculated $^3He$ upper limit and one imprecise measurement of $^{21}Ne$ (Sec. 2.7).

This direct comparison of an elemental ratio, in this case $^4He/^{20}Ne$, with nucleosynthetic theory is similar to that carried out by Heck et al. (2007) for $^4He/^{22}Ne$ ratios measured in their study of He and Ne in presolar SiC grains. Elemental composition is revealed here because the anomalous IDP gases apparently contain little or no interference from normal solar-system components. In this respect they differ from the first detections of presolar noble gases in meteorites (Reynolds & Turner 1964; Black & Pepin 1969; Black 1972). There the isotopic signatures were diluted by mixing with other meteoritic gases, and the pure compositions were determined only when their carrier grains —nanodiamonds for Xe-HL, graphite and silicon carbide for Ne-E— were chemically or physically separated from the bulk meteorites (Anders & Zinner 1993; Nittler 2003; Zinner 2004). This is not the case for the nova-linked IDPs, even though the much larger sizes of the ~8–15 μm fragments analyzed in this study compared to the ~0.4–1.4 μm grains observed in nova outflows (Gehrz 2008) suggest that minerals from other sources are dominant constituents of the anomalous IDPs. Agreements of measured He and Ne isotope ratios with theoretically derived ejecta compositions indicate that extraneous non-nova materials in the samples contribute little if at all to the gas inventories.

A striking feature of Fig. 1 is the non-correlation of the $^{20}$Ne and $^{22}$Ne concentrations in the anomalous IDPs. With one exception, $^{22}$Ne is approximately uniform in the 9 samples, but $^{20}$Ne varies by a factor of ~7 and $^{20}$Ne/$^{22}$Ne ratios by a factor of ~6. Noble gases incorporated into minerals by mechanisms such as irradiation, solution, or physical trapping closely reflect the isotopic compositions of their sources. An example is the normal IDP group in Fig. 1 where the correlated $^{20}$Ne and $^{22}$Ne concentrations yield $^{20}$Ne/$^{22}$Ne ratios consistent with that in the irradiating SW. Uncorrelated $^{20}$Ne and $^{22}$Ne in the anomalous group requires separate sources for the two isotopes. We propose that most of the $^{22}$Ne derives from decay of $^{22}$Na condensed in grains during their growth in the expanding nova outflow, and the non-condensable $^{20}$Ne and $^{4}$He reflect shock or stellar wind emplacement into the grains from an ambient reservoir strongly depleted in $^{22}$Ne and $^{3}$He (Sec. 4.4).

*4.3. Alternative Stellar Sources*

Other sources for the anomalous He and Ne appear unlikely. Nucleosynthesis in AGB stars (Heck et al. 2007) and CO novae (José et al. 2004, Heck et al. 2009) generates $^{20}Ne/\Sigma^{22}Ne$ ratios that are far too low. Comparison of the total yield ranges for type II supernovae (Rauscher et al. 2002) in Fig. 6 with the anomalous IDP values in Fig. 7 shows that the $^{4}He/^{20}Ne$, $^{20}Ne/\Sigma^{22}Ne$, and $^{20}Ne/^{21}Ne$ SN II ratios are significantly offset from the IDP data, in directions also indicating underproduction of $^{20}Ne$. Total yields, however, reflect complete mixing of nucleosynthesis products generated in separate supernova shells. Infrared spectroscopic observations of a supernova remnant (Cas A) suggest that shell products may be largely unmixed in the ejecta (Rho et al. 2008), so grains condensing within individual shells could have acquired locally synthesized gases. However, for the case of ejecta from a 15 solar mass supernova (Rauscher et al. 2002*,* data accessed at www.nucleosynthesis.org), Fig. 8 indicates that He and Ne isotopes produced in any of the separate SN II shells cannot reproduce the full set of anomalous IDP ratios, nor can those generated by mixing adjacent shells. Integrated Ne yields in the Ni shell come closest in Fig. 8 to matching the average IDP Ne ratios, and $^{3}He/^{4}He$ (not shown) is $<<10^{-4}$. However $^{4}He/^{20}Ne$ in the shell is too high by a factor of $\sim 10^{4}$. For the mixing alternative, the best possibility appears to be a combination of products synthesized in the Ni through O/C shells, collectively ~10% of total ejecta mass, which yields $^{3}He/^{4}He$ $<<10^{-4}$, $^{4}He/^{20}Ne$ reasonably close to the average IDP ratio, and $^{20}Ne/\Sigma^{22}Ne$ (~35) near the lower end of the measured range. But this mixing scenario cannot account for $^{20}Ne/\Sigma^{22}Ne$ ratios >35 in 7 of the 9 anomalous IDPs (Fig. 3), nor for a calculated $^{20}Ne/^{21}Ne$ lower than the measured ratio by $\sim 6\sigma$.

*4.4. Grain Condensation and Gas Emplacement in Nova Outflows*

Irrespective of white dwarf class, ONe or CO, grains begin to nucleate and grow to maximum size in cooling nova outflows within ~80 days or less after the eruptions (Gehrz 1988). While dominant mineralogies depend in part on nova class, individual novae are capable of producing a suite of grain types including C, SiC, silicates and hydrocarbons (José et al. 2004; Gehrz 2008). Infrared emission from grains evolves to maximum luminosity at an average of roughly 90 days when temperature are ~900-1000K (Gehrz 2008, Table 8.1). Most rock-forming elements will have condensed by this time (Lodders 2003).

Of primary interest, in the context of the $^{20}$Ne and $^{22}$Ne measurements, is the time scale for the condensation of Na-bearing minerals. In the modeled ONe ejecta composition chosen here to compare with the noble gas results (Tables 6 & 7), the ratio of total Na to the major silicate-forming elements (O+Mg+Al+Si) is ~10% by mass (Starrfield et al. 2009). A possible Na mineral could be albite ($NaAlSi_3O_8$), which condenses at ~960K from a solar composition gas (Lodders 2003). Grain temperatures near this are reached approximately 90 days after the explosion. At this time only 6.4% of the 2.6-year $^{22}$Na generated in the outburst will have decayed to $^{22}$Ne during its pre-condensation residence in the nova outflow. The resulting $^{20}$Ne/$^{22}$Ne ratio in the ejecta gas phase at the time of Na condensation is ~800 from data in Table 6. Subsequent implantations of this ambient Ne into grains, in amounts required by measured $^{20}$Ne in the anomalous IDPs (Fig. 1), contribute <10% of their average $^{22}$Ne concentration. The $^{22}$Ne inventories remain dominated by in-situ decay of $^{22}$Na in grains even if Na precipitation occurs substantially later. For example, at 240 days when grain temperatures are estimated to be ~200-400 K (Gehrz et al. 1986), ambient $^{20}$Ne/$^{22}$Ne would be ~320 and the average implanted $^{22}$Ne contribution <25%. This sequestering of $^{22}$Na in grains as dust condenses and cools in the nova outflow, together with destruction of pre-existing $^{3}$He and $^{22}$Ne in the TNR, results in a gas-phase He and Ne composition strongly depleted in $^{3}$He and $^{22}$Ne.

Laboratory experiments have demonstrated that shocks are effective in implanting ambient gases into solids (e.g., Wiens & Pepin 1988). Shocks, albeit of a different type, should be common in nova outburst environments. It is well known that Rayleigh-Taylor instability causes outflow fragmentation in stellar winds (Shore 2007, section 9.6.1). The manifestation of this fragmentation is seen in the clumpiness evident in optical images of the shells of old novae such as HR Del (Harman & O'Brien 2003). Dust grains in nova ejecta are most likely to grow within these clumps where they are shielded from ionizing radiation (Gehrz 2008), an effect graphically illustrated by separation of the dust formation zone and the ionization front in the wind of RY Scuti (Smith et al. 2001). The clumps will be subjected to strong radiative shock heating (Shore 2007, section 4.5.4) by the hard radiation field of the white dwarf central engine, which has a measured surface temperature as high as $8 \times 10^5$ K by the time the grains have matured (Petz et al. 2005). Such shocks could lead to stochastic implantation of ambient Ne and He into grains embedded in the outflow clumps, in amounts governed by local shock conditions.

*4.5. Speculations on Gas Carriers*

The physical, chemical, and mineralogic character of the anomalous IDP gas carriers has not been determined. The expectation, however, is that ONe novae with low C/O ratios will condense silicates, as observed in novae QU Vul 1984#2 (Gehrz et al. 1986), V2362 Cyg (Helton et al. 2011), and V1065 Cen (Helton et al. 2010). Size estimates of grains emitting silicate spectral features are in the range ~0.4–1.4μm, possibly eroded to smaller sizes before ejection into the interstellar medium (Gehrz 2008). An extensive SIMS-SEM search for presolar grains in IDPs from the Grigg-Skjellerup collection (GSC) (Busemann et al. 2009) revealed presolar silicates with high abundances, up to >1% by weight (wt.%), and sizes (~0.34μm) notably similar to the spectroscopically identified nova silicates. With abundance of 1 wt.%, roughly 100 of the GSC presolar grains would be present in the average anomalous IDP. If they were the carriers of the nova-linked noble gases, the required He loading would be on the order of 1 $cm^3$STP $g^{-1}$ (~1 He atom per 700 silicate atoms), comparable to levels seen in heavily SW-irradiated IDPs (e.g., Fig. 4 & Pepin et al. 2000).

However the role of the GSC presolar silicates as the actual nova gas carriers is questionable. Isotopic ratio ranges for C, N, O, Mg and Si calculated from nuclear modeling of neon nova ejecta compositions are shown in Fig. 9 (Starrfield et al. 2009; José et al. 2004). The GSC silicate grains exhibit the expected $^{17}$O enrichments, but their Si isotopic compositions differ from the Fig. 9 predicted pattern (Busemann et al. 2009). It may be that the carriers have a distinctly different mineralogy. Carbonaceous matter, inferred from spectral observations of the Ne-rich novae V2361 Cyg and V2362 Cyg (Helton et al. 2011), has long been known to be a principal host of noble gases in meteorites (e.g., Frick & Pepin 1981). It is relatively abundant in the GSC IDPs (Busemann et al. 2009; Nakamura-Messenger et al. 2008, 2010).

The actual nova gas carriers should display signatures of the non-solar isotopic distributions in C, N, O, Mg and Si predicted by nuclear modeling (Fig. 9). Dilution with normal solar system matter would mute these signatures, as is the case for the Amari et al. (2001) analyses of SiC grains of putative origin in ONe novae, but element-to-element patterns could be preserved. If such anomalies were found, their hosts would help to establish the nature of the gas-carrying grains.

## 5. CONCLUSIONS

The isotopic distributions of Ne and He measured in the anomalous IDPs are unique compared to solar-system compositions. Their similarities to isotope ratios theoretically predicted for nucleosynthesis in neon nova TNRs are evidence that dust produced in classical nova explosions, the most common type of stellar outburst in the Galaxy, found its way into the early solar system. Among alternative stellar sources for the noble gases in these IDPs, AGB stars and CO novae are ruled out, and mixed or unmixed ejecta from type II supernovae appear unlikely.

We summarize the central theoretical and observational features of a neon nova outburst as they relate to the anomalous IDP measurements. [1] Nuclear fusion reactions in the TNR powering the outburst annihilate preexisting $^{22}$Ne and $^{3}$He; $^{20}$Ne/$^{22}$Ne is driven to $>10^6$, $^{3}$He/$^{4}$He to $<10^{-8}$. The TNR generates radiogenic $^{22}$Na (half-life = 2.60 yr, mean life $\tau$ = 3.75 yr). [2] Condensations of ≈ 0.4-1.4μm silicate dust grains are detected by IR spectroscopy in ONe nova outflows ≲80 days after the TNR. Grain IR luminosities peak at ~90 days when grain temperatures T are ~1000K. Carbonaceous condensates are also observed in some neon novae ejecta. Grains tend to condense in clumps formed by Rayleigh-Taylor instabilities in the outflow, shielded from ionizing radiation from the hot WD central engine. Grain clumps are likely to experience strong radiative shocks. [3] With no source other than $^{22}$Na decay, $^{22}\mathrm{Ne} = (^{22}\mathrm{Na})_o(1 - e^{-t/\tau})$ in the outflow gas at time t after the TNR, where $(^{22}\mathrm{Na})_o$ is the initial TNR-generated abundance. At t ~90 days, $^{22}\mathrm{Ne} \simeq 0.06\,(^{22}\mathrm{Na})_o$ and gas phase $^{20}$Ne/$^{22}$Ne ≃ 800.

A model based on these nova characteristics can account for He and Ne distributions in the anomalous IDPs. [1] Na-bearing minerals (e.g., albite) condense at ~90 days, T ~1000K, removing $^{22}$Na from the expanding gas and freezing gas-phase $^{20}$Ne/$^{22}$Ne at ≃800. [2] In situ decay of precipitated $^{22}$Na in grains generates the $^{22}$Ne inventories of the noble gas carriers. Carrier grains are distributed roughly uniformly throughout the larger volumes of the anomalous IDPs, resulting in approximately constant concentrations of $^{22}$Ne/g-IDP. [3] Variable amounts of ambient $^{22}$Ne-poor Ne and $^{3}$He-free He are implanted into carrier grains by WD-powered radiative shocks on grain clumps. At $^{20}$Ne/$^{22}$Ne = 800 in the gas phase, too little $^{22}$Ne is implanted to significantly perturb Na-derived $^{22}$Ne grain inventories. The spread in implanted amounts of $^{20}$Ne reflects stochastic variations in shock-driven fluxes and energies of the ambient gas.

Despite the good agreement of theoretical calculations and astrophysical observations with the IDP measurements, arguments for a neon nova provenance of the anomalous noble gases will not be decisive until their carrier materials are identified. The carriers could be anything observed to condense in classical nova outflows. For the ONe class, silicates are the most obvious candidates but not the only ones. Whatever their nature, the most compelling marker of nova origin would be the presence, however subdued, of the predicted non-solar isotopic signatures in other elements shown in Fig. 9.

Given the rarity of presolar grains in meteorites thought to possibly originate in novae, the presence of nine samples in the Grigg-Skjellerup and Tempel-Tuttle comet collections containing noble gas carriers attributable to this source is surprising. These targeted collections, however, are uniquely distinguished from earlier random IDP samplings. There is strong evidence, particularly in the agreement, discussed in Sec. 3.2.1, of estimated space exposure histories of GSC anomalous IDPs with that of the GS dust stream, that they contain the first materials directly traceable to cometary sources. Unusual features have been identified in some of these IDPs, including a new mineral. The noble gas data additionally indicate the presence of presolar matter from a neon nova outburst.

We thank D. J. Schlutter for assistance with mass spectrometry measurements, and M. Zolensky and staff of the NASA Cosmic Dust Curatorial Facility for IDP sample selection and preparation. S. Messenger and S. Shore are thanked for discussions and comments on the manuscript. This research was supported in Physics at Minnesota by the NASA LARS Program, in Astronomy at Minnesota by NSF, NASA, and the US Air Force, and at Arizona State by NSF and NASA.

## REFERENCES

Amari, S., Gao, X., Nittler, L. R., Zinner, E., José, J., Hernanz, M., & Lewis, R. S. 2001, ApJ, 551, 1065

Anders, E., & Zinner, E. 1993, Meteoritics, 28, 490

Becker, R. H. 1998, Lunar Planet. Sci. Conf., XXIX, abstract #1329

Black, D. C., & Pepin, R. O. 1969, Earth Planet. Sci. Lett., 6, 395

Black, D. C. 1972, Geochim. Cosmochim. Acta, 36, 377

Bode, M. F., & Evans, A., ed. 2008, *Classical Novae* (Cambridge: Cambridge Univ. Press)

Busemann, H., Nguyen, A. N., Cody, G. D., Hoppe, P., Kilcoyne, A. L. David, Stroud, R. M., Zega, T. J., & Nittler, L. R. 2009, Earth Planet. Sci. Lett., 288, 44

Eberhardt, P., Eugster, O., & Marti, K. 1965, Z. Naturforsch., 20a, 623

Frick, U., & Pepin, R. O. 1981, Earth Planet. Sci. Lett., 56, 45

Gehrz, R. D., Grasdalen, G. L., & Hackwell, J. A. 1985, ApJ, 298, L47

Gehrz, R. D., Grasdalen, G. L., Greenhouse, M., Hackwell, J. A., Hayward, T., & Bentley, A. F. 1986, ApJ, 308, L63

Gehrz, R. D. 1988, Ann. Rev. Astron. Astrophys., 26, 377

Gehrz, R. D., Truran, J. W., Williams, R. E., & Starrfield, S. 1998, PASP, 110, 3

Gehrz, R. D. 2008, in Classical Novae, ed. M. F. Bode, & A. Evans (Cambridge: Cambridge Univ. Press), 167

Harman, D. J., & O'Brien, T. J. 2003, MNRAS, 344, 1219

Heber, V. S., Wieler, R., Baur, H., Olinger, C., Friedmann, T. A., & Burnett, D. S. 2009, Geochim. Cosmochim. Acta, 73, 7414

Heck, P. R., Marhas, K. K., Hoppe, P., Gallino, R., Baur, H., & Wieler, R. 2007, ApJ, 656, 1208

Heck, P. R., Amari, S., Hoppe, P., Baur, H., Lewis, R. S., & Wieler, R. 2009, ApJ, 701, 1415

Helton, L. A., et al. 2010, ApJ, 140,1347

Helton, L. A., Gehrz, R. D., Woodward, C. E., & Evans, A. 2011, in PAHs and the Universe, ed. C. Joblin, & A. G. G. M. Tielens (EAS Publications Series), in press

José, J., Hernanz, M., Amari, S., Lodders, K., & Zinner, E. 2004, ApJ, 612, 414

Joswiak, D. J., Brownlee, D. E., Pepin, R. O., & Schlutter, D. J. 2000, Lunar Planet. Sci. Conf., XXXI, abstract #1500

Kehm, K., Flynn, G. J., Sutton, S. R., & Hohenberg, C. M. 2002, Meteorit. Planet. Sci., 37, 1323

Kehm, K., Flynn, G. J., & Hohenberg, C. M. 2006, Meteorit. Planet. Sci., 41, 1199

Lodders, K. 2003, ApJ, 591, 1220

Love, S. G., & Brownlee, D. E. 1994, Meteoritics, 29, 69

Mabry, J. C., et al. 2007, Lunar and Planetary Science, XXXVIII, abstract #2412

Mahaffy, P. R., Donahue, T. M., Atreya, S. K., Owen, T. C., & Niemann, H. B. 1998, in ISSI Space Sci. Ser., Vol. 4, Primordial Nuclei and their Galactic Evolution, ed. R. von Steiger et al. (Norwell, MA: Kluwer Acad.), 251

Mattauch, J., & Herzog, R. 1934, Zeitschrift für Physik, 89, 786

Messenger, S. 2002, Meteorit. Planet. Sci., 37, 1491

Nakamura-Messenger, K., Keller, L. P., Messenger, S., Clemett, S. J., Zolensky, M. E., Palma., R. L., Pepin, R. O., & Wirick, S. 2008, Meteorit. Planet. Sci., 43, A111

Nakamura-Messenger, K., et al. 2010, Am. Mineralogist, 95, 221

Nier, A. O., & Schlutter, D. J. 1985, Rev. Sci. Instr., 56, 214

Nier, A. O., & Schlutter, D. J. 1990, Meteoritics, 25, 263

Nittler, L. R. 2003, EPSL, 209, 259

Nittler, L. R., & Hoppe, P. 2005, ApJ, 631, L89

Nittler, L. R., Alexander, C. M. O'D., & Nguyen, A. N. 2006, Meteorit. Planet. Sci., 41 Supp., A134

Owen, T., Mahaffy, P. R., Niemann, H. B., Atreya, S., & Wong, M. 2001, Ap J, 553, L77

Palma, R. L., Pepin, R. O., & Schlutter, D. 2005, Meteorit. Planet. Sci., 40, A120

Pepin, R. O., Becker, R. H., & Schlutter, D. J. 1999, Geochim. Cosmochim. Acta, 63, 2145

Pepin, R. O., Palma, R. L., & Schlutter, D. J. 2000, Meteorit. Planet. Sci., 35, 495

Pepin, R. O., Palma, R. L., & Schlutter, D. J. 2001, Meteorit. Planet. Sci., 36, 1515

Petz, A., Hauschildt, P. H., Ness, J.-U., & Starrfield, S. 2005, in ASP Conf. Vol. 330, The Astrophysics of Cataclysmic Variables and Related Objects, ed. J.-M. Hameury, & J.-P. Lasota (San Francisco, CA: ASP), 299

Rauscher, T., Heger, S., Hoffman, R. D., & Woosley, S. E. 2002, ApJ, 576, 323

Reisenfeld, D. B., et al. 2007, Space Sci. Rev., 130, 79

Reynolds, J. H., & Turner, G. 1964, J. Geophys. Res., 69, 3263

Rho, J., Kozasa, T., Reach, W. T., Smith, J. D., Rudnick, L., DeLaney, T., Ennis, J. A., Gomez, H., & Tappe, A. 2008, ApJ, 673, 271

Shore, S. N. 2007, Astrophysical Hydrodynamics: An Introduction, Second Edition (Weinheim: WILEY-VCH Verlag GmbH & Co KGaA)

Smith, N., Gehrz, R. D., & Goss, W. M. 2001, AJ, 122, 2700

Starrfield, S., Iliadis, C., & Hix, W. R. 2008, in Classical Novae, ed. M. F. Bode, & A. Evans (Cambridge: Cambridge Univ. Press), 77

Starrfield, S., Iliadis, C., Hix, W. R., Timmes, F. X., & Sparks, W. M. 2009, ApJ, 692, 1532

Thomas, K. L., et al. 1995, Geochim. Cosmochim. Acta, 59, 2797

Wiens, R. C., & Pepin, R. O. 1988, Geochim. Cosmochim. Acta., 52, 295

Zinner, E. K. 2004, in Meteorites, Comets, and Planets, Vol. 1, Treatise in Geochemistry, ed. A. M. Davis, H. D. Holland, & K. K. Turekian (1st ed; Oxford: Elsevier), 17

Zolensky, M. 2004, Availability of particles from collection surfaces exposed in the stratosphere during the 2002 Leonid Shower (comet 55P/Tempel-Tuttle), and Earth's predicted crossing of the dust steam from comet Grigg-Skjellerup in 2003 (Houston, TX: JSC), http://curator.jsc.nasa.gov/dust/print_leonid.cfm

**Table 1**
Helium and neon abundances and isotope ratios in individual Suite 1 IDPs.

| Sample<br>Date§, mass | ID¶ | $^{4}He$ [$10^{-11}$ $cm^{3}$STP] | $^{20}Ne$ [$10^{-12}$ $cm^{3}$STP] | $^{4}He/^{20}Ne$ | $^{3}He/^{4}He$ [x $10^{-4}$] | $^{20}Ne/^{22}Ne$ | $^{21}Ne/^{22}Ne$ [x $10^{-2}$] | $^{20}Ne/^{21}Ne$ |
|---|---|---|---|---|---|---|---|---|
| L2054 A3*3‡<br>4/28, 0.30 ng | 1 | nd | 4.00<br>±0.06 | nd: $^{3}He$, $^{4}He$ | | 152<br>±56 | nd: $^{21}Ne$ | |
| L2054 A4*4‡<br>4/28, 1.9 ng | 2 | 0.651<br>±0.011 | 3.73<br>±0.19 | 1.74<br>±0.09 | nd: $^{3}He$ | 50.6<br>±9.6 | 3.8<br>±3.2 | 1420<br>(710-9800) |
| L2054 A1*1†<br>5/4, 1.6 ng | 3 | 7.70<br>±0.10 | 2.57<br>±0.09 | 30.0<br>±1.1 | 2.84<br>±0.23 | 14.2<br>±2.1 | 7.1<br>±2.5 | 200<br>±52 |
| L2055 A2*4†<br>5/7-12, 2.1 ng | 4 | 36.1<br>±0.4 | 2.98<br>±0.29 | 121<br>±12 | 4.63<br>±0.15 | 13.5<br>±2.9 | 6.3<br>±3.3 | 215<br>±120 |
| L2055 A1*1‡<br>5/13, 0.30 ng | 5 | 0.257<br>±0.008 | 2.91<br>±0.10 | 0.883<br>±0.041 | nd: $^{3}He$ | 71<br>±21 | nd: $^{21}Ne$ | |
| U2108 A1*1‡<br>5/25, 0.36 ng | 6 | 0.214<br>±0.012 | 2.39<br>±0.14 | 0.895<br>±0.073 | nd: $^{3}He$ | 47<br>±18 | nd: $^{21}Ne$ | |
| L2055 A5*7†<br>5/26-27, 0.45 ng | 7 | 4.28<br>±0.09 | 1.21<br>±0.11 | 35.4<br>±3.3 | 3.72<br>±0.57 | 11.5<br>±2.8 | nd: $^{21}Ne$ | |
| L2055 A4*5†<br>5/28-6/2, 1.2 ng | 8 | 26.6<br>±0.4 | 1.57<br>±0.09 | 169<br>±10 | 2.56<br>±0.13 | 16.0<br>±4.8 | nd: $^{21}Ne$ | |
| L2055 A4*6‡<br>6/2, 0.30 ng | 9 | nd | 0.667<br>±0.047 | nd: $^{3}He$, $^{4}He$ | | 25<br>±15 | nd: $^{21}Ne$ | |
| L2055 A5*8‡<br>6/7, 0.28 ng | 10 | 1.10<br>±0.01 | 0.742<br>±0.068 | 14.8<br>±1.4 | nd: $^{3}He$ | 28<br>±20 | nd: $^{21}Ne$ | |
| U2121 A3*3†<br>6/14, 0.45 ng | 11 | 1.27<br>±0.04 | 1.50<br>±0.10 | 8.47<br>±0.62 | 3.4<br>±1.0 | 12.4<br>±1.8 | 5.8<br>±1.6 | 213<br>±54 |
| U2121 A1*1‡<br>6/15, 0.55 ng | 12 | 0.069<br>±0.004 | 2.31<br>±0.09 | 0.299<br>±0.021 | nd: $^{3}He$ | 49<br>±17 | nd: $^{21}Ne$ | |
| U2108 A3*3‡<br>6/15, 0.43 ng | 13 | 2.06<br>±0.06 | 5.47<br>±0.15 | 3.77<br>±0.15 | nd: $^{3}He$ | 145<br>±43 | nd: $^{21}Ne$ | |
| U2121 A2*2‡<br>6/16, 0.55 ng | 14 | 0.887<br>±0.023 | 4.09<br>±0.07 | 2.17<br>±0.07 | nd: $^{3}He$ | 66<br>±12 | nd: $^{21}Ne$ | |
| U2108 A2*2†<br>6/16, 0.61 ng | 15 | 6.19<br>±0.08 | 3.33<br>±0.13 | 18.59<br>±0.76 | 4.09<br>±0.39 | 13.2<br>±1.4 | 3.44<br>±0.82 | 383<br>±87 |

Notes. IDP: Interplanetary Dust Particle, nd: not detected. Uncertainties are ± 1σ analytic errors.

¶ IDP identifier in Fig. 1.

§ Date is analysis date (month/day).

‡ Low-He Anomalous IDP. Masses from particle dimensions and assumed cometary IDP density of 1.1 g $cm^{-3}$; uncertainties ≈ ± 70%.

† High-He Normal IDP. Masses from particle dimensions and assumed asteroidal IDP density of 2.5 g $cm^{-3}$; uncertainties ≈ ± 70% .

**Table 2**

Individual step temperatures and He and Ne data for Suite 1 IDPs degassed and analyzed in multiple heating steps.

| IDP | Table 1 Identifier | Mass [ng] | Run date [2004] | Temperature [°C] | $^{4}$He [$10^{-12}$ cm$^{3}$STP] | $^{20}$Ne [$10^{-12}$ cm$^{3}$STP] | $^{3}$He/$^{4}$He [x $10^{-4}$] | $^{20}$Ne/$^{22}$Ne | $^{21}$Ne/$^{22}$Ne [x $10^{-2}$] |
|---|---|---|---|---|---|---|---|---|---|
| L2054 A3*3 | | 0.30 | 4/28 | 1380 | nd: $^{4}$He | 4.00(06) | nd: $^{3}$He | 152(56) | nd: $^{21}$Ne |
| | | | 4/28 | 1410 | nd | nd | — | — | — |
| | 1 | | | *Totals* | *nd: $^{4}$He* | *4.00(06)* | *nd: $^{3}$He* | *152(56)* | *nd: $^{21}$Ne* |
| L2054 A1*1 | | 1.6 | 5/4 | 750 | 73.7(10) | 1.39(07) | 2.84(23) | 11.8(14) | 8.6(23) |
| | | | 5/4 | 1400 | 3.34(05) | 1.18(06) | nd: $^{3}$He | 18.9(39) | 4.3(26) |
| | 3 | | | *Totals* | *77.0(10)* | *2.57(09)* | *2.84(23)* | *14.2(21)* | *7.1(25)* |
| L2055 A2*4 | | 2.1 | 5/7 | 790 | 36.8(16) | 1.02(16) | 5.63(52) | 16.1(47) | 6.3(33) |
| | | | 5/7 | 1110 | 39.1(15) | 1.30(19) | 3.19(33) | nd: $^{22}$Ne | — |
| | | | 5/10 | 1370 | 35.6(14) | nd | 4.36(60) | — | — |
| | | | 5/10 | 1410 | 57.1(17) | nd | 4.57(36) | — | — |
| | | | 5/11 | 1410 | 49.5(16) | 0.05(02) | 5.98(57) | nd: $^{22}$Ne | — |
| | | | 5/11 | 1410 | 58.0(15) | nd | 4.81(29) | — | — |
| | | | 5/12 | 1425 | 83.4(18) | 0.61(16) | 4.11(26) | 10.6(33) | nd: $^{21}$Ne |
| | | | 5/12 | 1425 | 1.76(09) | nd | nd: $^{3}$He | — | — |
| | 4 | | | *Totals* | *361(04)* | *2.98(29)* | *4.63(15)* | *13.5(29)* | *6.3(33)* |
| U2108 A1*1 | | 0.36 | 5/24 | 520 | 33.5(13) | 0.22(03) | 3.00(32) | 8.6(53) | nd: $^{21}$Ne |
| | | | 5/24 | 720 | 36.4(14) | 0.25(03) | 1.79(34) | 9.9(27) | nd: $^{21}$Ne |
| | | | 5/25 | 870 | 25.2(11) | 0.45(04) | 1.86(58) | 11.9(35) | 8.0(55) |
| | | | 5/25 | 1430 | 2.14(12) | 2.39(14) | nd: $^{3}$He | 47(18) | nd: $^{21}$Ne |
| | † | | | *Totals, 520-870 °C* | *95.1(22)* | *0.92(06)* | *2.23(23)* | *10.4(24)* | *8.0(55)* |
| | 6 | | | *Totals, 1430 °C* | *2.14(12)* | *2.39(14)* | *nd: $^{3}$He* | 47 (18) | *nd: $^{21}$Ne* |

**Table 2 (continued)**

| IDP | Table 1 Identifier | Mass [ng] | Run date [2004] | Temperature [°C] | $^{4}$He [$10^{-12}$ cm$^{3}$STP] | $^{20}$Ne [$10^{-12}$ cm$^{3}$STP] | $^{3}$He/$^{4}$He [x $10^{-4}$] | $^{20}$Ne/$^{22}$Ne | $^{21}$Ne/$^{22}$Ne [x $10^{-2}$] |
|---|---|---|---|---|---|---|---|---|---|
| L2055 A5*7 | | 0.45 | 5/26 | 310 | 19.9(06) | 0.10(04) | 1.95(82) | nd: $^{22}$Ne | — |
| | | | 5/27 | 890 | 22.9(07) | 0.71(08) | 5.25(78) | 12.8(36) | nd: $^{21}$Ne |
| | | | 5/27 | 1410 | nd | 0.40(06) | nd: $^{3}$He | 9.8(41) | nd: $^{21}$Ne |
| | 7 | | | *Totals* | *42.8(09)* | *1.21(11)* | *3.72(57)* | *11.5(28)* | *nd: $^{21}$Ne* |
| L2055 A4*5 | | 1.2 | 5/28 | 490 | 91.2(22) | 0.24(02) | 2.50(27) | nd: $^{22}$Ne | — |
| | | | 5/28 | 640 | 106(03) | 0.47(03) | 2.75(13) | 26 (11) | nd: $^{21}$Ne |
| | | | 6/1 | 720 | 47.6(17) | 0.09(02) | 2.17(27) | 5.6(71) | nd: $^{21}$Ne |
| | | | 6/1 | 1080 | 20.8(08) | 0.77(08) | 2.75(73) | 15.9(32) | nd: $^{21}$Ne |
| | | | 6/2 | 1400 | nd | nd | — | — | — |
| | 8 | | | *Totals* | *266(04)* | *1.57(09)* | *2.56(13)* | *16.0(48)* | *nd: $^{21}$Ne* |
| U2121 A3*3 | | 0.45 | 6/14 | 1350 | 10.5(04) | 1.50(10) | 2.61(60) | 12.4(18) | 5.8(16) |
| | | | 6/14 | 1420 | 2.18(11) | nd | 7.4(52) | — | — |
| | 11 | | | *Totals* | *12.7(04)* | *1.50(10)* | *3.4(10)* | *12.4(18)* | *5.8(16)* |

† Not plotted in Fig. 1

**Table 3**
Helium and neon abundances and isotope ratios in individual Suite 2 IDPs.

| Sample<br>Mass† | $^{4}$He<br>[$10^{-11}$ cm$^{3}$STP] | $^{20}$Ne<br>[$10^{-12}$ cm$^{3}$STP] | $^{4}$He/$^{20}$Ne | $^{3}$He/$^{4}$He<br>[x $10^{-4}$] | $^{20}$Ne/$^{22}$Ne | $^{21}$Ne/$^{22}$Ne<br>[x $10^{-2}$] | $^{20}$Ne/$^{21}$Ne |
|---|---|---|---|---|---|---|---|
| L2054 L4*1<br>0.04 ng | 2.73<br>±0.03 | 0.082<br>±0.006 | 332<br>±25 | 2.97<br>±0.35 | | nd: $^{21}$Ne, $^{22}$Ne | |
| L2054 L3*2<br>0.91 ng | 3.29<br>±0.09 | nd: $^{20}$Ne | | 4.27<br>±0.33 | | nd: $^{21}$Ne, $^{22}$Ne | |
| L2054 L1*3<br>0.08 ng | 0.998<br>±0.024 | nd: $^{20}$Ne | | 2.9<br>±1.2 | | nd: $^{21}$Ne, $^{22}$Ne | |
| L2054 L2*4<br>0.16 ng | 0.903<br>±0.021 | 0.299<br>±0.016 | 30.2<br>±1.8 | 4.9<br>±1.3 | nd: $^{22}$Ne | | 88<br>±55 |
| L2055 I1*1‡<br>0.20 ng | 0.117<br>±0.003 | 0.035<br>±0.004 | 33.4<br>±3.9 | nd: $^{3}$He | | nd: $^{21}$Ne, $^{22}$Ne | |
| L2055 I2*6‡<br>0.16 ng | 0.107<br>±0.003 | 0.097<br>±0.005 | 11.0<br>±0.6 | nd: $^{3}$He | | nd: $^{21}$Ne, $^{22}$Ne | |
| L2055 I3*7<br>0.08 ng | 1.85<br>±0.04 | 0.195<br>±0.018 | 94.9<br>±9.0 | 1.60<br>±0.59 | | nd: $^{21}$Ne, $^{22}$Ne | |
| L2055 L1*7<br>0.40 ng | 25.6<br>±0.3 | 0.927<br>±0.072 | 276<br>±22 | 2.99<br>±0.13 | 18<br>±11 | 0.12<br>±0.10 | 152<br>±83 |
| U2121 E1*1<br>0.29 ng | 6.22<br>±0.05 | 0.478<br>±0.021 | 130.1<br>±5.8 | 2.54<br>±0.28 | nd: $^{22}$Ne | | 290<br>±260 |
| U2121 E2*2<br>1.2 ng | 3.94<br>±0.13 | nd: $^{20}$Ne | | 3.24<br>±0.22 | | nd: $^{21}$Ne, $^{22}$Ne | |

Notes. IDP: Interplanetary Dust Particle, nd: not detected. Uncertainties are ± 1σ analytic errors.
† Masses estimated from particle dimensions and assumed asteroidal IDP density of 2.5 g cm$^{-3}$; uncertainty ≈ ± 70%.
‡ Possible low-He Anomalous IDP.

**Table 4**

Solar wind isotope ratios in Genesis collector materials.

| | $^{4}He/^{20}Ne$ | $^{3}He/^{4}He$ [x $10^{-4}$] | $^{20}Ne/^{22}Ne$ | $^{21}Ne/^{22}Ne$ [x $10^{-2}$] | $^{20}Ne/^{21}Ne$ |
|---|---|---|---|---|---|
| Solar Wind[1] | 656 ± 5 | 4.64 ± 0.09 | 13.78 ± 0.03 | 3.29 ± 0.01 | 418.8 ±1.6 |
| Solar Wind[2] | 643 | 4.34 ±0.02 | 13.97 ± 0.03 | 3.46 ±0.03 | 403.8 ±2.4 |
| Average Solar Wind | 650 ± 7 | 4.49 ±0.15 | 13.88 ± 0.10 | 3.38 ±0.09 | 411.3 ±7.5 |

References: [1]Heber et al. 2009, [2]Mabry et al. 2007

**Table 5**

Concentrations of $^{20}Ne$ and $^{22}Ne$ plotted in Fig. 1, calculated from Table 1 data.

| Sample | Figure 1 Identifier | Mass [ng] | Collection | Type | $^{20}Ne^{\dagger}$ [$10^{-3}$ $cm^3$ STP/g] | $^{22}Ne^{\dagger}$ |
|---|---|---|---|---|---|---|
| L2054 A3*3 | 1 | 0.30 | Grigg-Skjellerup | Anomalous | 13.47(20) | 0.088(33) |
| L2054 A4*4 | 2 | 1.9 | Grigg-Skjellerup | Anomalous | 1.94(10) | 0.038(07) |
| L2054 A1*1 | 3 | 1.6 | Grigg-Skjellerup | Normal | 1.58(05) | 0.111(16) |
| L2055 A2*4 | 4 | 2.1 | Grigg-Skjellerup | Normal | 1.40(14) | 0.104(22) |
| L2055 A1*1 | 5 | 0.30 | Grigg-Skjellerup | Anomalous | 9.80(33) | 0.138(41) |
| U2108 A1*1 | 6 | 0.36 | Tempel-Tuttle | Anomalous | 6.58(39) | 0.140(53) |
| L2055 A5*7 | 7 | 0.45 | Grigg-Skjellerup | Normal | 2.69(24) | 0.233(57) |
| L2055 A4*5 | 8 | 1.2 | Grigg-Skjellerup | Normal | 1.26(08) | 0.079(24) |
| L2055 A4*6 | 9 | 0.30 | Grigg-Skjellerup | Anomalous | 2.25(16) | 0.088(50) |
| L2055 A5*8 | 10 | 0.28 | Grigg-Skjellerup | Anomalous | 2.70(25) | 0.095(66) |
| U2121 A3*3 | 11 | 0.45 | Grigg-Skjellerup | Normal | 3.33(22) | 0.269(39) |
| U2121 A1*1 | 12 | 0.55 | Grigg-Skjellerup | Anomalous | 4.20(16) | 0.086(30) |
| U2108 A3*3 | 13 | 0.43 | Tempel-Tuttle | Anomalous | 12.75(38) | 0.088(26) |
| U2121 A2*2 | 14 | 0.55 | Grigg-Skjellerup | Anomalous | 7.44(13) | 0.113(21) |
| U2108 A2*2 | 15 | 0.61 | Tempel-Tuttle | Normal | 5.44(22) | 0.412(46) |

† Numbers in parentheses are analytic uncertainties in the last two digits of the listed concentrations.

**Table 6**
Isotope concentrations in white dwarf (WD) neon nova ejecta. Data from sequence I2005A for 1.25 solar mass ONe WD (Starrfield et al. 2009), converted from g $g^{-1}$ ejecta to $cm^3$ STP $g^{-1}$ ejecta.

| $^4He$ | $^3He$ | $^{20}Ne$ | $^{21}Ne$ | $^{22}Ne$ | $^{22}Na$ |
|---|---|---|---|---|---|
| | | ($cm^3$ STP $g^{-1}$ ejecta) | | | |
| $8.95 \times 10^2$ | $5.07 \times 10^{-6}$ | $2.35 \times 10^2$ | $1.07 \times 10^{-1}$ | $2.03 \times 10^{-4}$ | $4.58 \times 10^0$ |

**Table 7**
Helium and neon concentrations and isotope ratios in IDPs and Neon Nova ejecta.

| | $^4He$ ($10^{-3}$ $cm^3$STP $g^{-1}$) | $^{20}Ne$ ($10^{-3}$ $cm^3$STP $g^{-1}$) | $^4He/^{20}Ne$ | $^3He/^4He$ (x $10^{-4}$) | $^{20}Ne/^{22}Ne$ | $^{21}Ne/^{22}Ne$ (x $10^{-2}$) | $^{20}Ne/^{21}Ne$ |
|---|---|---|---|---|---|---|---|
| Normal IDPs* | 197<br>(28 – 780) | 2.4<br>(1.3– 5.4) | 111<br>(8 – 332) | 3.33 ± 0.91<br>(1.6 – 4.9) | 13.5 ± 1.5<br>(11.5 – 15.9) | 5.4 ± 1.5<br>(3.4 – 7.1) | 262 ± 84<br>(200 – 383) |
| Anomalous IDPs* | 17.6<br>(1.3 – 48) | 6.8<br>(1.9 – 13.5) | 3.5<br>(0.3 – 14.8) | < 1 | 70<br>(25 – 152) | 3.8 ± 3.2‡ | 1420‡<br>(710 – 9800) |
| Neon Nova ejecta† | $8.95 \times 10^5$ | $2.35 \times 10^5$ | 3.8 | $5.7 \times 10^{-5}$ | 51§ | 2.33§ | 2200 |

Notes. IDP: Interplanetary Dust Particle.
* 1st lines are averages, 2nd lines are ranges of data in Table 1 for anomalous IDPs, in Tables 1 and 3 for normal IDPs.
† Calculated from Table 6 data.
‡ $^{21}Ne$ measured in only 1 of the 9 anomalous IDPs (see Table 1). Uncertainties are ±1σ analytic errors.
§ $^{22}Ne = \Sigma^{22}Ne = (^{22}Ne + ^{22}Na)$ in Neon Nova ejecta.

## FIGURE CAPTIONS

**Figure 1.** Concentrations (Table 5) of $^{20}$Ne and $^{22}$Ne in 15 IDPs from the Grigg-Skjellerup (GSC) and Tempel-Tuttle (TTC) cometary collections. One cm$^3$STP contains 2.69 x 10$^{19}$ atoms. Data point numbers identify the correspondingly numbered individual samples in Table 5. Average densities of particles thought to derive from asteroids and comets are 2.5 and 1.1 g cm$^{-3}$ respectively (Joswiak et al. 2000, Sec. 3.3). Neon concentrations per gram are measured isotopic abundances divided by masses calculated from particle dimensions, assuming asteroidal density for the normal IDPs and cometary density for the anomalous group. The plotted ± 1$\sigma$ error bars reflect errors in measured abundances; they do not include the larger uncertainties in estimates of particle masses (Sec. 3.3). The dashed line indicates the average $^{22}$Ne concentration of 0.097 ± 0.031 x 10$^{-3}$ cm$^3$STP g$^{-1}$ in the 9 anomalous IDPs; 8 of the 9 fall on the line within their errors, indicating essentially constant $^{22}$Ne concentrations. The one outlier with low $^{22}$Ne (#2) is the largest IDP in the anomalous group, with a mass (at 1.1 g cm$^{-3}$) ~5-fold greater than the average of the others. Its structure is unknown, but it would also lie on the line if it were a highly porous particle with a density of ~0.5 g cm$^{-3}$.

**Figure 2.** Comparison of $^{22}$Ne/$^{20}$Ne and $^{4}$He/$^{20}$Ne ratios in anomalous and normal IDPs. $^{4}$He was not detected (nd) in two of the anomalous samples. Plotted uncertainties are ± 1$\sigma$. Data sources, keyed to symbol colors: Red and blue, this work, Suite 1, Table 1; Green, Nier & Schlutter (1990); Orange, Pepin et al. (2000); Yellow, unpublished data from the Minnesota laboratory; White, Kehm et al. (2002, 2006). Solar wind composition from Table 4.

**Figure 3.** Fig. 2 data replotted on a $^{20}Ne/^{22}Ne$ vs. linear $^{4}He/^{20}Ne$ diagram. Indicated uncertainties for $^{20}Ne/^{22}Ne$ in the anomalous IDPs are $\pm 1\sigma$ (Table 1); they inherit the analytic errors involved in measurements of small amounts of $^{22}Ne$ (Sec. 2.6) and are consequently substantial. SW-$^{20}Ne/^{22}Ne$ ratio from Table 4; SW-$^{4}He/^{20}Ne$ is offscale to the right at ~650. Data sources as in Fig. 2.

**Figure 4.** Comparison of He isotope ratios and $^{4}He$ concentrations in anomalous and normal IDPs. Error (not shown) in the anomalous IDP $^{4}He$ concentration due to uncertainties in particle masses is ~ ±70% of the plotted value (Sec. 3.3). Symbols for the normal IDPs represent averages of individual samples analyzed in each data source, and error bars their $\pm 1\sigma$ standard deviations in $^{3}He/^{4}He$; adjacent numbers give the number of samples in each group. $^{3}He$ was not detected (nd) in any of the 9 anomalous IDPs. Data sources as in Fig. 2, with addition of: Light blue, Suite 2, Table 3; Light orange, Pepin et al. (2001). Solar wind $^{3}He/^{4}He$ from Table 4.

**Figure 5.** Measured He isotopic abundances in anomalous and normal IDPs. The normal group distribution is well fit by the power law $^{3}He = 5.90 \times 10^{-5} [^{4}He]^{0.926}$. Individual anomalous IDPs other than U2108 A3*3 are also plotted, with the assumption that the $^{3}He/^{4}He$ ratio of $<10^{-4}$ determined for U2108 A3*3 (Sec. 2.7) applies to them as well. One additional sample is offscale with $^{4}He < 10^{-12}$ $cm^{3}STP$; two others contained no detectable $^{4}He$ (Table 1). Data sources as in Fig. 4.

**Figure 6.** Modeled He and Ne isotope ratios in neon nova ejecta. Solar wind ratios (Table 4) are plotted for comparison. Nova ejecta compositions (✻) are for a 1.25 solar mass ONe white dwarf

(WD) from Table 7. Vertical bars indicate the ranges of ratios generated in models of neon nova outbursts over a WD mass range of 1.00 to 1.35 solar masses with several different choices for initial compositions and nuclear reaction networks (José et al. 2004). $\Sigma^{22}$Ne represents the sum of $^{22}$Ne and the 2.6-year $^{22}$Na synthesized in the TNR. Also shown for comparison are compositional ranges for total ejecta from models of nucleosynthesis in type II supernovae (SN II) of 15 and 25 solar masses (Rauscher et al. 2002).

**Figure 7.** Measured He and Ne isotope ratios in the normal and anomalous IDPs analyzed in this work compared with SW compositions and TNR modeling results. IDP data from Table 7 (nd: not detected), SW from Table 4. Data points are average ratios in the normal and anomalous groups; thin vertical bars indicate minimum and maximum measured values. Neon nova modeling results from Fig. 6.

**Figure 8.** Compositions of individual shells in a post-explosion 15 $M_{\odot}$ SN II supernova. Shell data from Rauscher et al. (2002), model s15a28c, accessed at www.nucleosynthesis.org. Shell nomenclature from Heck et al. (2009), neon nova ranges and anomalous IDP compositions from Fig. 7. Indicated isotope ratios are ratios of integrated isotopic abundances generated in all zones within each shell, with radiogenic $^{22}$Na production included in $\Sigma^{22}$Ne. “Total”, also shown in Fig. 6, represents summation of synthesized products over all shells plus ejected wind —i.e., completely mixed ejecta.

**Figure 9.** Predicted C, N, O, Mg and Si isotope ratios generated in neon nova TNRs. Neon nova (✻) from Starrfield et al. (2009), neon novae ranges from José et al. (2004). All ratios are normalized to solar system compositions, assumed to be terrestrial except for N where the Jupiter $^{15}N/^{14}N$ ratio of 2.3 x $10^{-3}$ (Owen et al. 2001) is used. Normalization to terrestrial N ($^{15}N/^{14}N$ = 3.68 x $10^{-3}$) would reduce the plotted $^{15}N/^{14}N$ enhancements by a factor of 1.6.

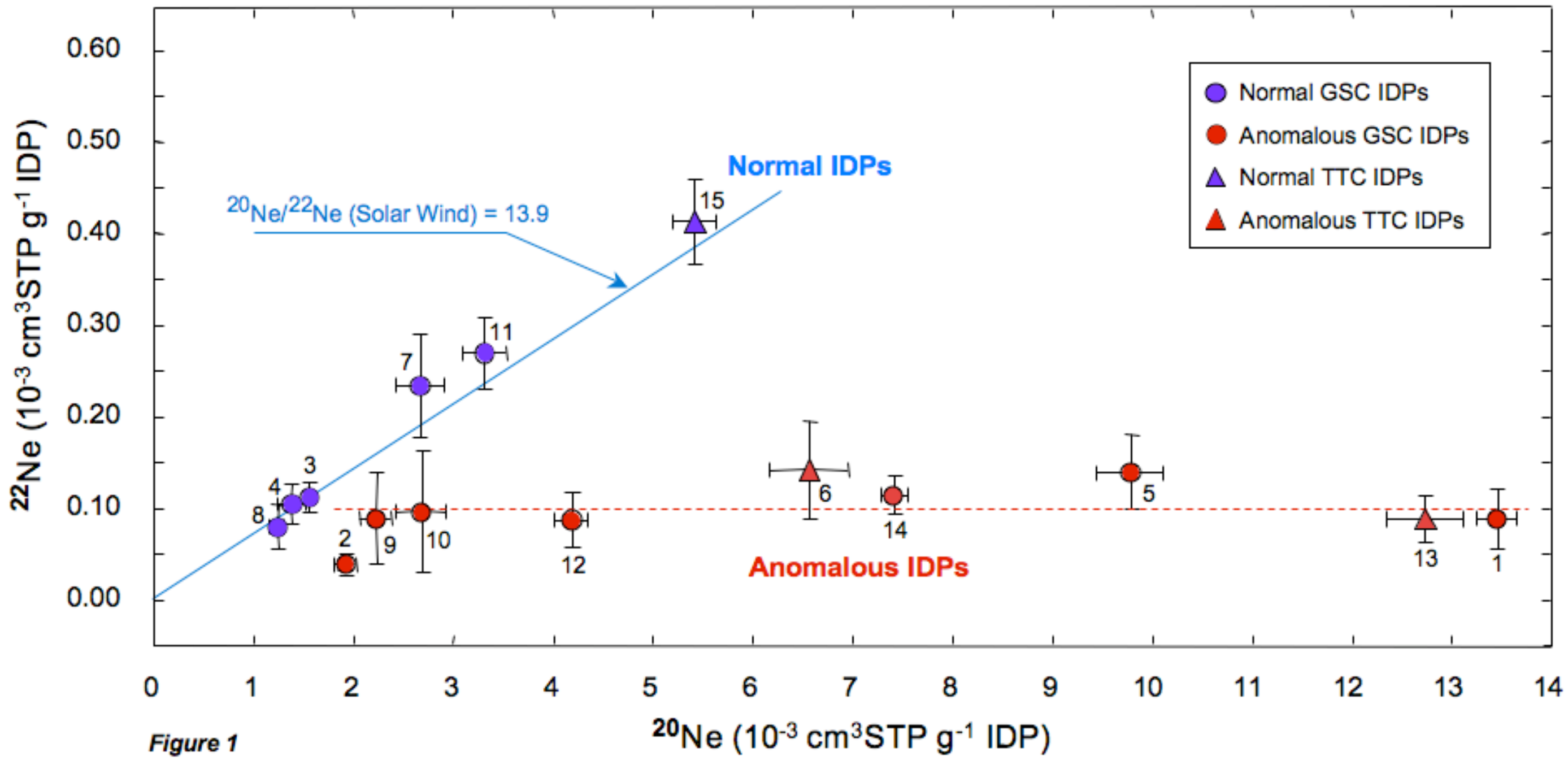

*Figure 1*

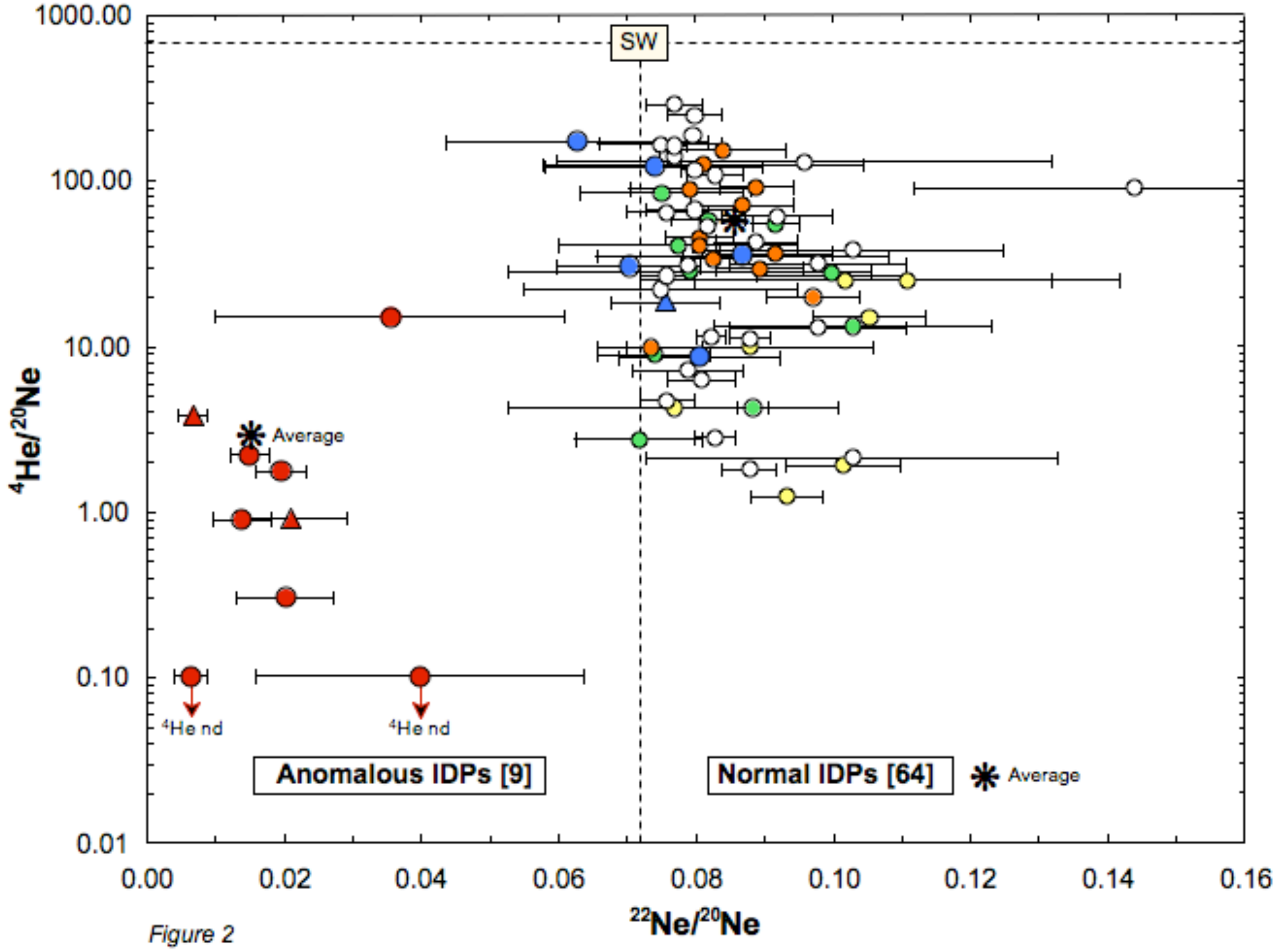

*Figure 2*

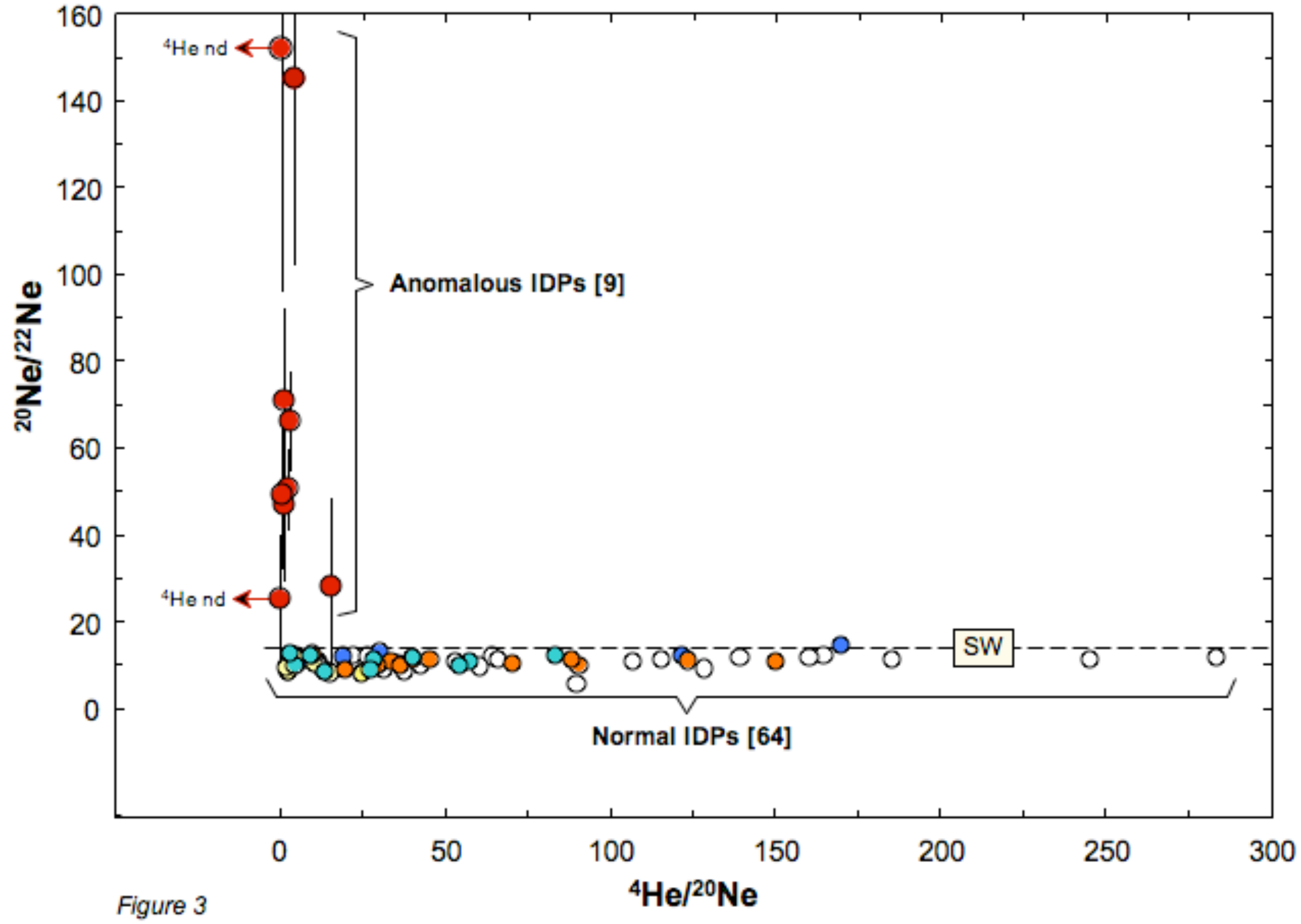

*Figure 3*

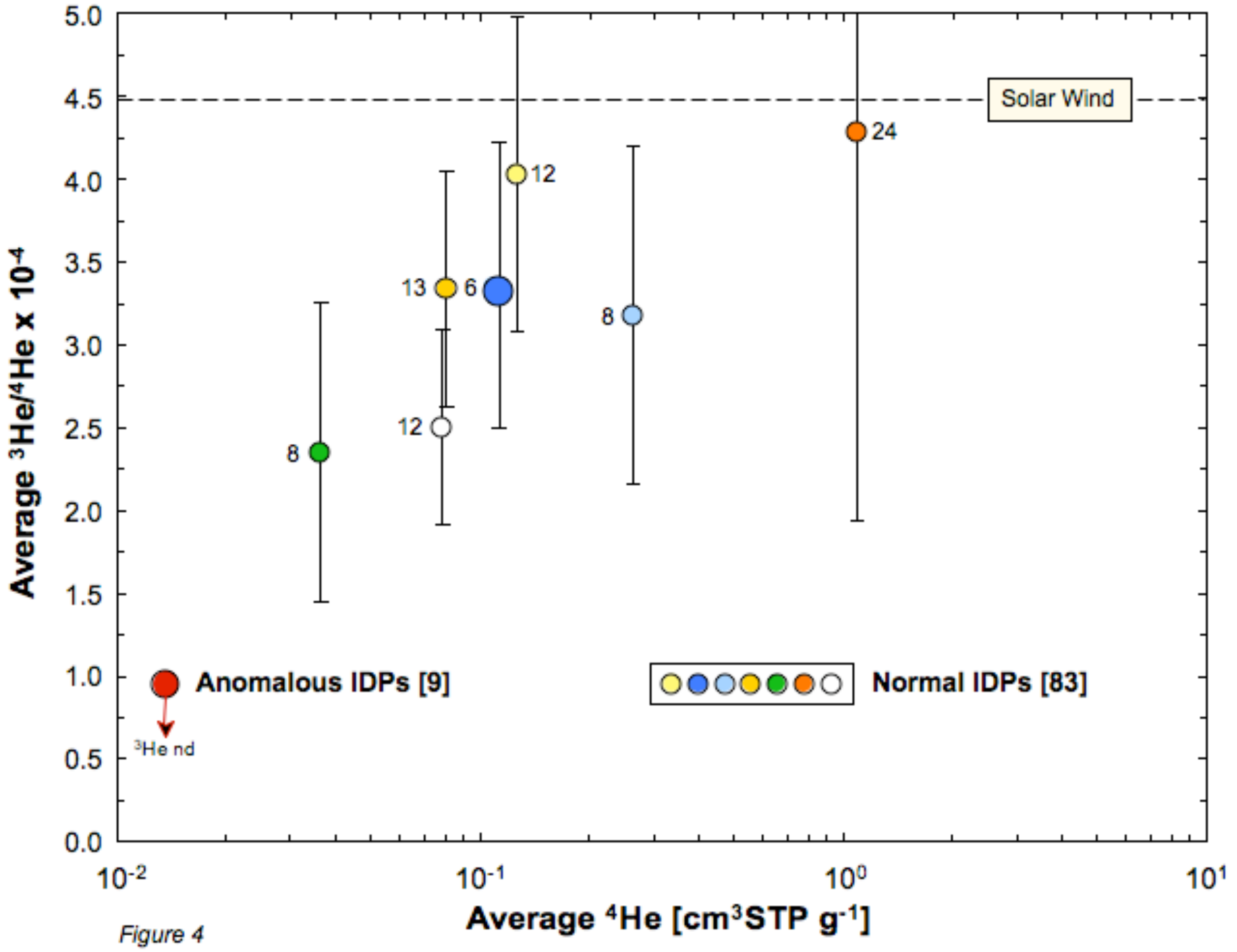

*Figure 4*

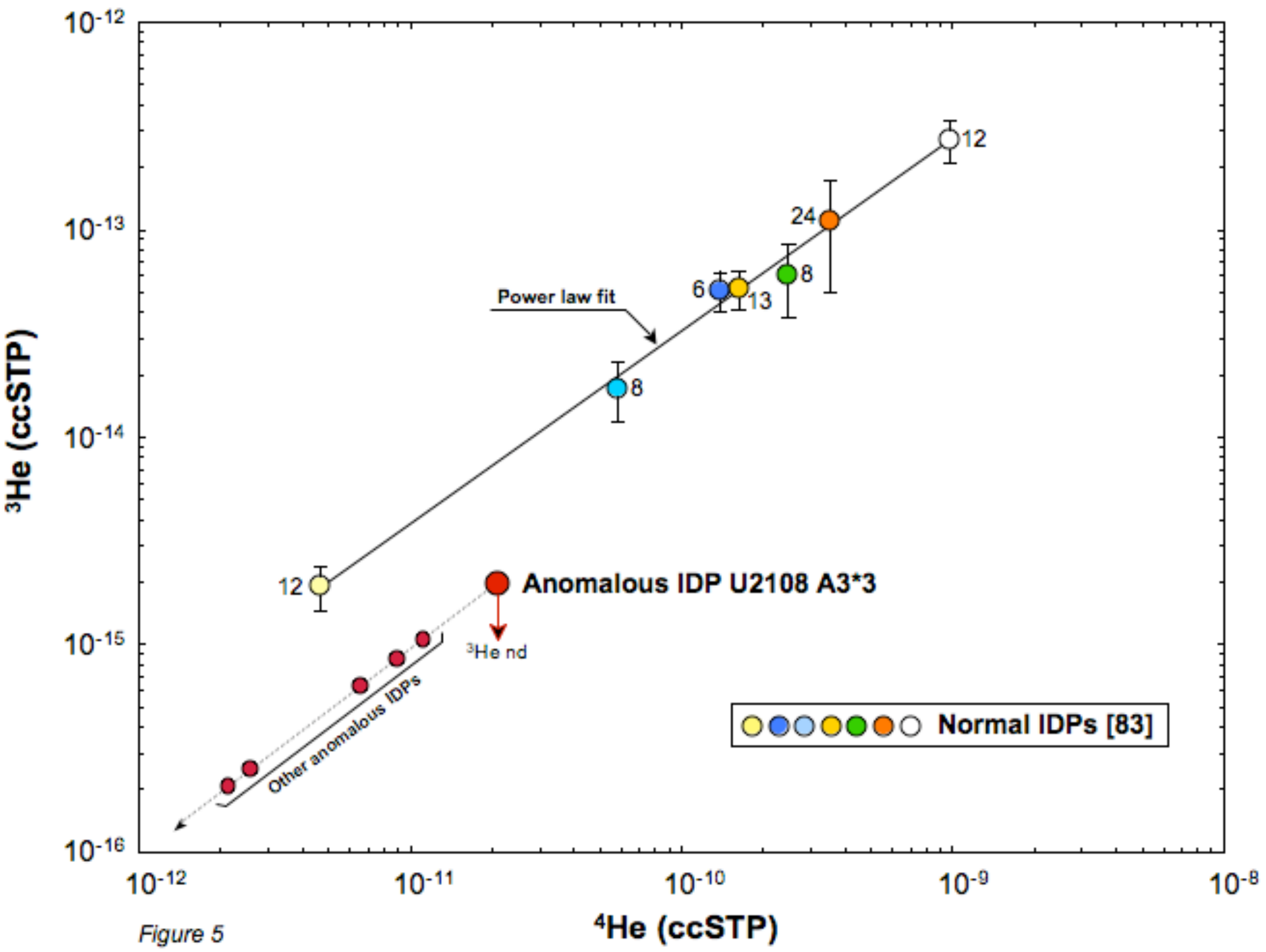

*Figure 5*

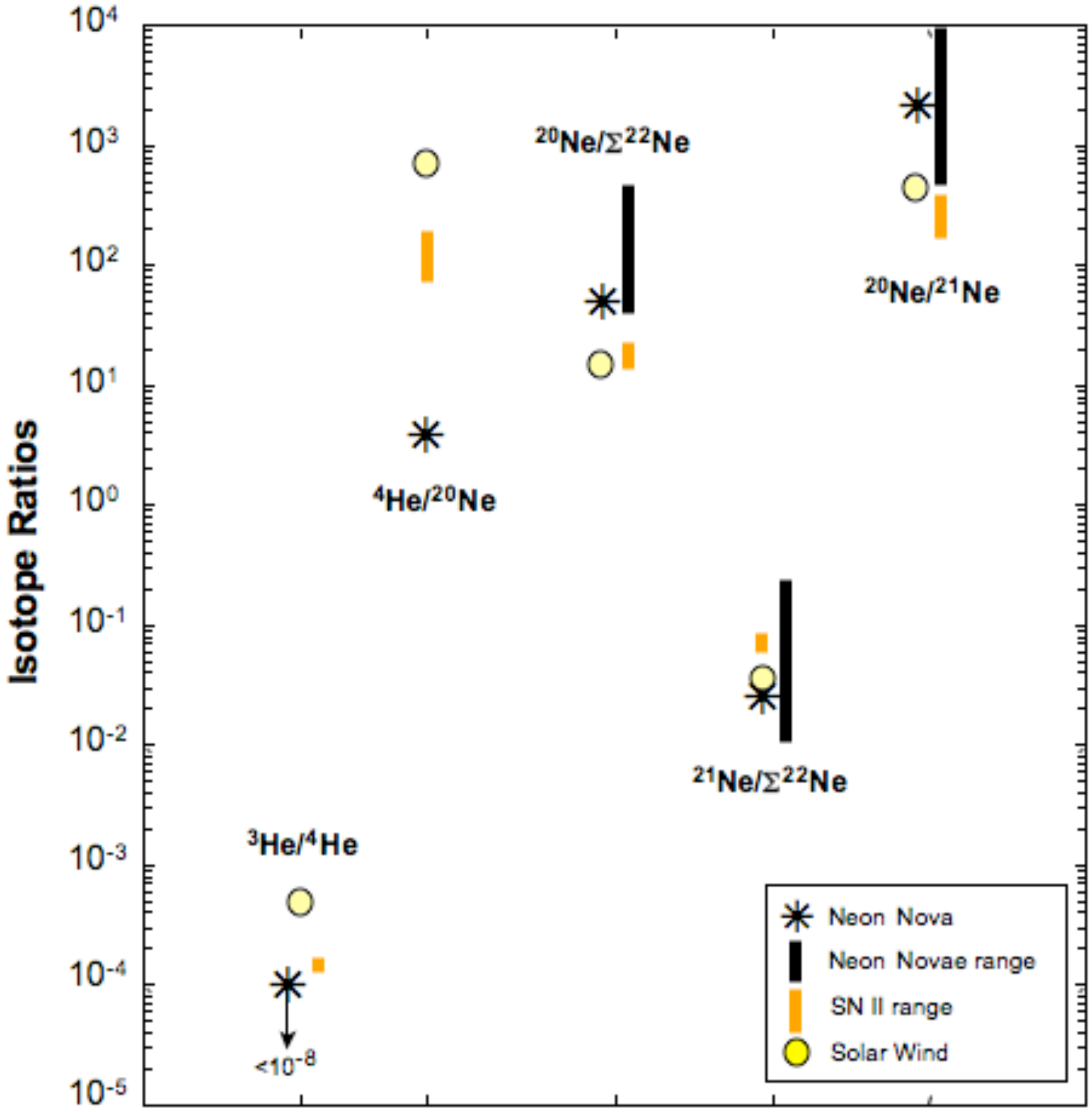

*Figure 6*

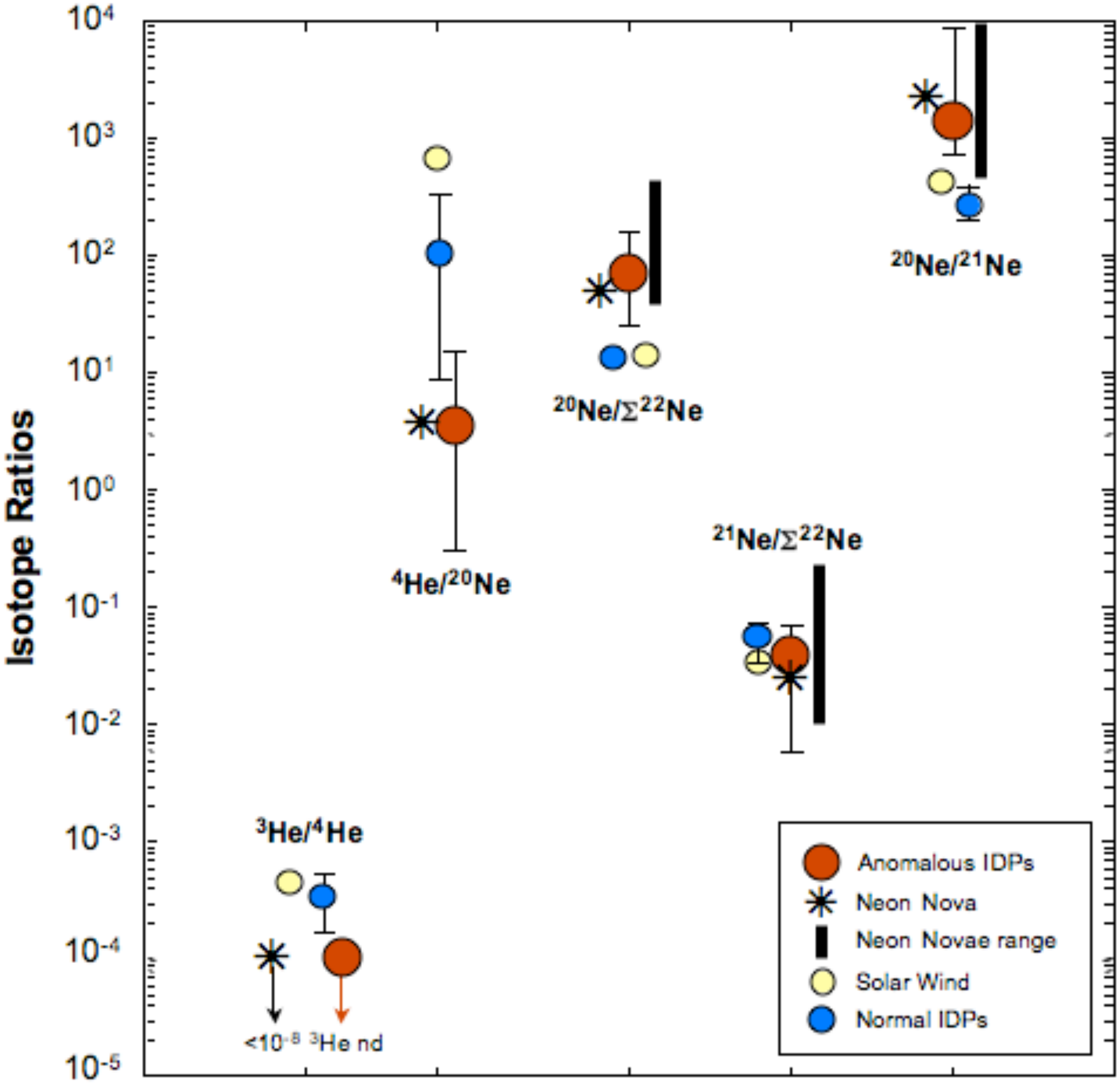

*Figure 7*

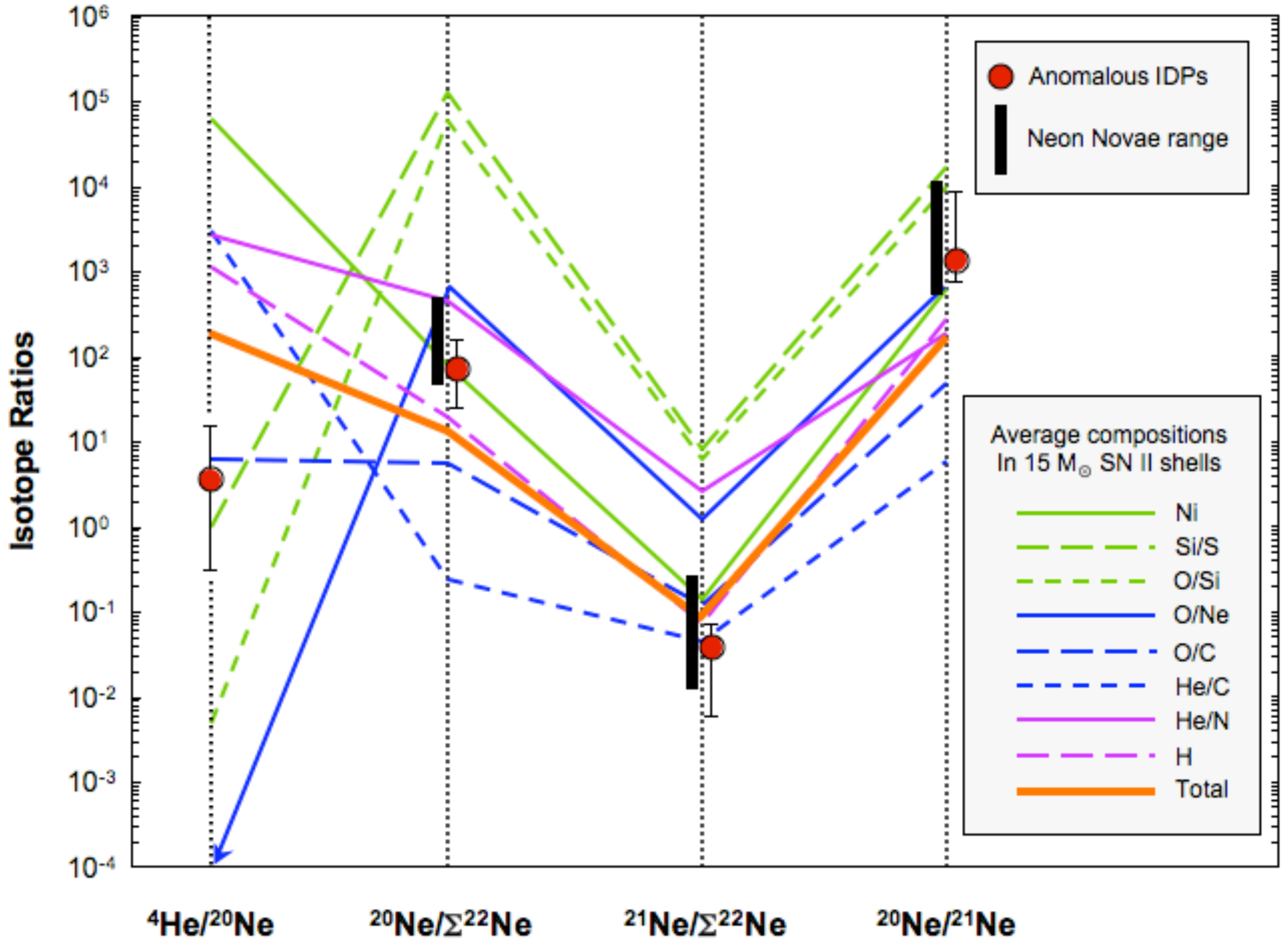
Anomalous IDPs
Neon Novae range
Average compositions In 15 M⊙ SN II shells
Ni
Si/S
O/Si
O/Ne
O/C
He/C
He/N
H
Total
Isotope Ratios
10^6
10^5
10^4
10^3
10^2
10^1
10^0
10^-1
10^-2
10^-3
10^-4
4He/20Ne
20Ne/Σ22Ne
21Ne/Σ22Ne
20Ne/21Ne

*Figure 8*

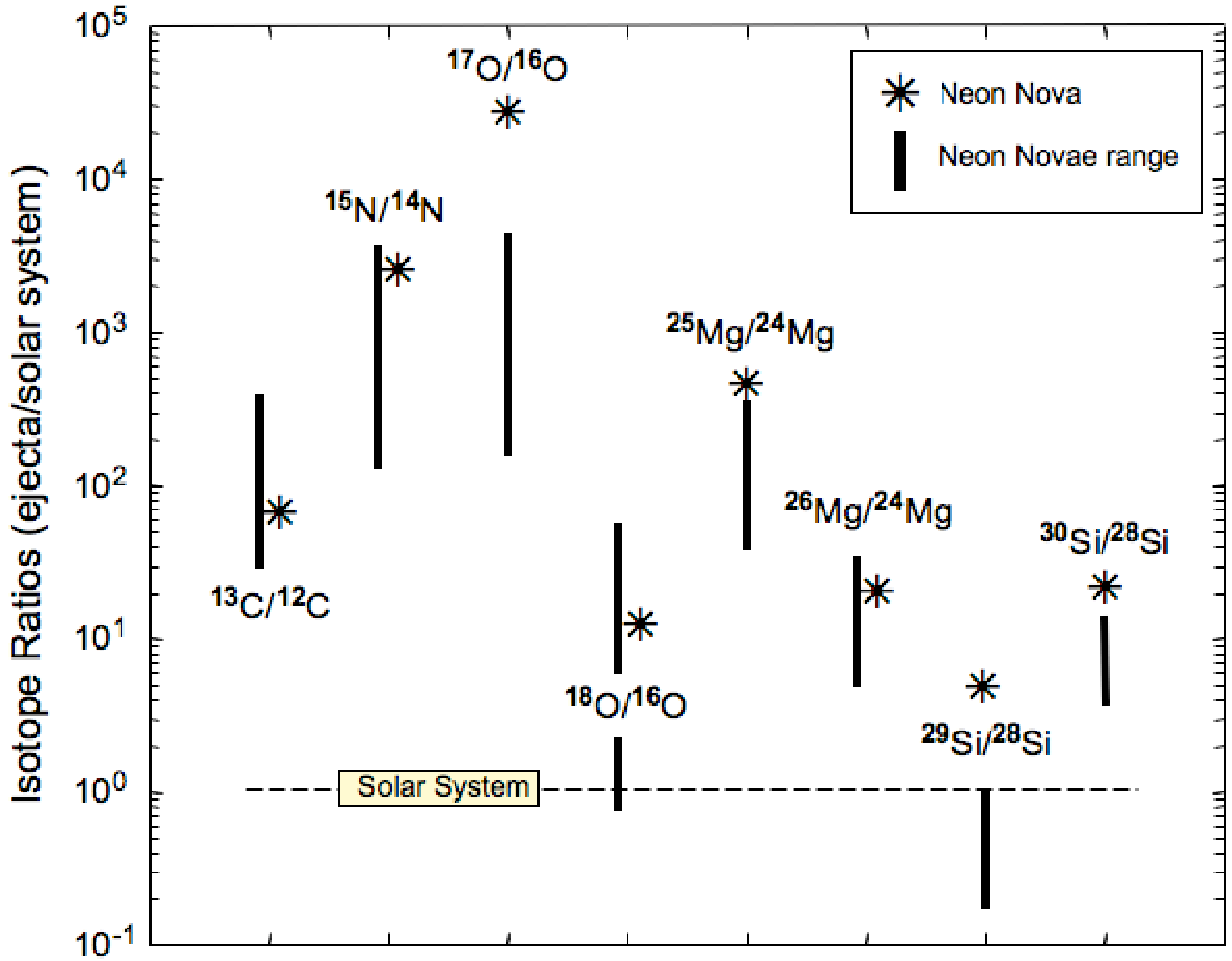

*Figure 9*